\documentclass{aa}

\usepackage{graphicx}
\usepackage{txfonts}
\usepackage{natbib}
\usepackage{hyperref}
\usepackage{multirow}
\usepackage{float}
\usepackage{placeins}
\usepackage{xcolor}

\graphicspath{{./figures/}}
\defcitealias{hamann2019}{H19}
\defcitealias{sander2019}{S19}
\defcitealias{sander2025}{S25}

\begin{document} 

   \title{Dynamically consistent analysis of Galactic WN4b stars}

   \subtitle{Revised parameters and insights on Wolf-Rayet mass-loss treatments}

   \author{%
   R.~R.~Lefever\inst{\ref{inst:ari}}
   \and
   A.~A.~C.~Sander\inst{\ref{inst:ari},\ref{inst:iwr}}
   \and
   M.~Bernini-Peron\inst{\ref{inst:ari}}
   \and
   G.~Gonz\'alez-Tor\`a\inst{\ref{inst:ari}}
   \and
   W.-R.~Hamann\inst{\ref{inst:potsdam}}
   \and
   J.~Josiek\inst{\ref{inst:ari}}
   \and
   V. Ramachandran\inst{\ref{inst:ari}}
   \and
   E.~C.~Sch\"osser\inst{\ref{inst:ari}}
   \and
   H.~Todt\inst{\ref{inst:potsdam}}
   }

   \institute{%
   {Zentrum f{\"u}r Astronomie der Universit{\"a}t Heidelberg, Astronomisches Rechen-Institut, M{\"o}nchhofstr. 12-14, 69120 Heidelberg\label{inst:ari}}\\
   \email{roel.lefever@uni-heidelberg.de}
   \and {Interdisziplin{\"a}res Zentrum f{\"u}r Wissenschaftliches Rechnen, Universit{\"a}t Heidelberg, Im Neuenheimer Feld 225, 69120 Heidelberg, Germany\label{inst:iwr}}
   \and {Institute of Physics and Astronomy, University of Potsdam, Karl-Liebknecht-Str. 24/25, 14476 Potsdam, Germany\label{inst:potsdam}}
   }

   \date{Received 24 November 2025; accepted 8 January 2026}

  \abstract
   {Many Wolf-Rayet (WR) stars have optically thick winds that essentially cloak the hydrostatic layers of the underlying star. 
   In these cases, traditional spectral analysis methods are plagued by degeneracies that make it difficult to constrain parameters such as the stellar radius and the deeper density and velocity structure of the atmosphere.}
   {Focussing on the regime of nitrogen-rich WN4-stars with strong emission lines, we employ hydrodynamically-consistent modelling using the PoWR$^\textsc{hd}$ code branch to perform a next-generation spectral analysis. 
   The inherent coupling of the stellar and wind parameters  enables us to break parameter degeneracies, constrain the wind structure, and get a mass estimate. 
   With this information, we can draw  evolutionary implications and test current mass-loss descriptions for WR stars.}
   {We selected a sample of six Galactic WN4b stars. Applying updated parallaxes from Gaia DR3 and calculating PoWR$^\textsc{hd}$ models that sufficiently resemble most of their spectral appearance, we obtain new values for the stellar and wind parameters of the WN4b sample.
   We compare our results to previous studies employing grid models with a prescribed $\beta=1$ velocity structure and cross-check our derived parameters with stellar structure predictions from GENEC and FRANEC evolution tracks.}
   {For all six targets, we obtain a narrow range of stellar temperatures $T_\ast\sim 140$ kK, in sharp contrast to previous grid-model analyses. We confirm the existence of WRs with luminosities as low as $\log L/L_\odot = 5.0$ and $M_\ast \approx 5\,M_\odot$.
   All derived velocity fields include a plateau-like feature at $\sim$85\% of the terminal velocity. Both the distance updates and the switch to dynamically-consistent atmospheres lead to substantial parameter adjustments compared to earlier grid-based studies. 
   A comparison of the derived mass-loss rates favours a different description for the WN4b sample than for WN2 stars analysed with the same methodology.}
   {WN4b winds are launched by the hot iron opacity bump, placing these hydrogen-free stars near or slightly hotter than the He zero age main sequence. 
   Similar to a recent analysis of WN2 stars, we thus solve the WR radius problem for the WN4b stars, but this conclusion cannot be extrapolated to regimes strongly affected by radiatively-driven turbulence. 
   Evolutionary models struggle to reproduce the empirical parameter combinations. 
   The observed stars typically require lower mass loss in the current WR stage than predicted, but would require further prior stripping in order to arrive at the observed stage. }


   \maketitle   
   
\section{Introduction}
   
\noindent In the domain of massive stars (stars with $M_\text{init}\gtrsim 8 M_\odot$), the Wolf-Rayet (WR) stars form a highly remarkable subgroup. 
Characterised by prominent emission lines in their spectra \citep[e.g.,][]{WolfRayet1867,Beals1940}, WR stars show strong stellar winds emerging from the proximity to the Eddington limit, i.e., a high ratio between luminosity and mass \citep[e.g.,][]{graefener2008,Sander+2020}. 
Most WR stars are found on the hottest and most luminous end of the Hertzsprung-Russell diagram (HRD). 
The combination of high temperatures and luminosities along with strong outflows causes these stars to be highly impactful: 
They serve as strong sources of ionising radiation \citep[e.g.,][]{smith2002,sander2022,sander2025}, contribute to local enrichment of their surroundings \citep[e.g.][]{maeder1983, dray2003, farmer2021} and inject large amounts of kinetic energy into their surroundings via their stellar winds \citep[e.g.][]{ramachandran2018}. 
These winds lead to mass-loss rates of $\dot{M} \sim 10^{-5} - 10^{-3}\,M_\odot\,\mathrm{yr}^{-1}$ and terminal velocities easily in the order of thousands of $\mathrm{km}\,\mathrm{s}^{-1}$ \citep[e.g.,][]{crowther2007}.
The powerful stellar wind heavily influences the WR-star spectra, up to the point where the spectrum is completely dominated by wind lines \citep{hamann1985,Davidson1987,hamann2004grids}. 
When the whole spectrum is formed in the rapidly expanding wind layers, the determination of the underlying stellar parameters gets particularly challenging.
The standard approach to analyse the spectra with an atmosphere model using prescribed wind velocity field is limited by significant degeneracies \citep[e.g.,][]{hamann2004grids,lefever2023}, in particular with respect to the (hydrostatic) stellar radius and its corresponding effective temperature. 
This limits the proper comparison with stellar structure and evolution models and has been termed the ``WR radius problem'' \citep[e.g.,][]{hillier1987,langer1988}.
This points towards the need for improvements in modelling techniques. 

Most known WR stars are highly evolved objects: About 90\% of all WR stars are in the helium burning stage (some may even be beyond) and thus are partially or completely depleted in hydrogen \citep[e.g.,][]{vdH2001,crowther2007,hainich2014lmc,shenar2019}.

These so-called ``classical'' WR stars (cWR) are further spectroscopically classified into WN stars with prominent nitrogen line presence, WC stars with strong carbon lines and WO stars with strong oxygen lines \citep{hiltner1966, smith1968}. 
Transition types between WN and WC as well as WN and WO have been discovered \citep{massey1989, morris1996,sander2025}, suggesting an evolutionary connection \citep{Gamow1943} and implying that at least some WR stars are capable of intrinsically removing a sufficient part of their outer envelope to alter their surface abundances \citep{conti1979, conti1983}.
Further temperature classification is done based on diagnostic line ratios within the three separate spectral classes (see \citealt{smith1996} for WN stars and \citealt{crowther1998} for WC and WO stars). 
A subgroup among the WN stars shows a notable presence of hydrogen at the surface and are classified as WNh stars.
These stars are compatible with core-hydrogen burning massive stars close to the Eddington limit \citep[e.g.,][]{dekoter1997,crowther2010,graefener2011}. 
However, the WNh classification can also be reached by incomplete stripping of the hydrogen envelope. 
This is more prominent at lower metallicity, in particular in the SMC where 11 out of 12 known WR stars are of WNh type, but presumably all of them are He-burning \citep{hainich2015smc,shenar2016}.

In all of these separate classifications, the stellar wind determines the appearance of the objects with significant consequences for the spectral analysis. 
Given the aforementioned ``WR radius problem'' for WR stars with dense winds obtained in traditional studies \citep[e.g.,][]{hamann2006}, the advent of dynamically-consistent modelling for WR stars \citep[e.g.,][]{graefener2005,Sander+2020} offers a new opportunity to resolve this long-standing issue. 

In this study, we focus on analysing Galactic dense-wind WR stars of the WN4 subtype, also labelled as WN4-s \citep{hamann1995} or WN4b \citep{smith1996} due to their strong, broad emission lines.
All of our six sample stars (WR\,1, WR\,6, WR\,7, WR\,18, WR\,37, and the WN/WC star WR\,58) are known to be in the regime of parameter degeneracy in prior studies \citep{hamann2006,sander2012}. 
In this work, we employ hydrodynamic consistency in our stellar atmosphere modelling \citep{graefener2005, sander2017} to obtain a physically more accurate information about their parameters, in particular with respect to their hydrostatic radii and wind onset.

Our paper is organised as follows: the methodology of our analysis, along with details on how dynamic consistency is achieved, is laid out in Sect. \ref{sec:methods}. 
In Sect. \ref{sec:results}, the key findings of our analysis are presented, with a special emphasis on the spectroscopic results while also being put in the context of stellar evolution. 
Interwoven with the scientific results, the comparison with previous analyses, focussing on differences due to the use of dynamically consistent wind modelling, is also provided in Sect. \ref{sec:results}.
The summary of this study's main findings and future prospects is given in Sect. \ref{sec:conclusions}.

\section{Methods and observational data}\label{sec:methods}

In order to accurately model WR-star atmospheres, a number of processes need to be simulated simultaneously. 
The strong departure from local thermodynamic equilibrium (LTE) requires an extensive calculation of the population numbers from the equations of statistical equilibrium. 
This non-LTE calculation is interdependent with the radiation field, which is obtained from solving the radiative transfer in a comoving-frame approach to account for the expanding atmosphere.
In addition, the (electron) temperature stratification needs to be determined, typically from radiative equilibrium, flux consistency, electron thermal balance, or a combination of these approaches. 
In this work, we use the PoWR stellar atmosphere code \citep{graefener2002, hamann2003temperature, sander2015consistent} to fulfill these tasks. 
Assuming spherical symmetry, PoWR in the standard version uses fixed input stellar and wind parameters ($M,\,T_\star,\,R_\star,\,L,\,\dot{M},\,\varv_\infty,\,\beta$) to compute the stratifications of hot-star atmospheres and their emergent spectra. 
The wind hydrodynamics are approximated by a fixed $\varv(r)$, usually using a $\beta$ velocity law, where $\varv(r) = \varv_\infty (1 - R_\star / r)^\beta$ \citep{castor1975radiation}. 
In the subsonic regime, this $\varv(r)$ is then smoothly attached to a hydrostatic solution \citep{sander2015consistent}.
In the case of WR stars, often $\beta=1$ is assumed \citep[e.g.][]{hamann1988, hillier1999}, in particular in the calculation of large model grids \citep[e.g.,][]{hamann2004grids, todt2015}, but this value can be insufficient to accurately reflect the wind hydrodynamics and can lead to troublesome degeneracies \citep[][]{graefener2005,Sander+2020,lefever2023}.

To overcome the limitations of the fixed-velocity approach, we therefore employ the PoWR$^\textsc{hd}$ branch \citet{sander2017}, which computes the velocity stratification from solving the hydrodynamic equation of motion
\begin{equation}\label{eq:force_balance}
    \varv \frac{\mathrm{d}\varv}{\mathrm{d} r} + \frac{GM}{r^2} = a_\mathrm{rad}(r) + a_\mathrm{press}(r)
\end{equation}
\noindent throughout the atmosphere consistently.
The left-hand side in Eq.\,\eqref{eq:force_balance} represents the repulsive forces, accelerating the wind (i.e., the gas pressure and the radiation forces), while on the right the attractive forces are written.
The acceleration due to gas and turbulence pressure is denoted as $a_\mathrm{press}$ while the radiative acceleration is termed $a_\mathrm{rad}$. 
Solving Eq.\,\eqref{eq:force_balance} not only determines $\varv(r)$ and thus the terminal velocity $\varv_\infty$, but the necessary boundary conditions (i.e., the conservation of the total optical depth in our case) also fix the mass-loss rate $\dot{M}$ to a unique value. 
For hot stars, these results for $\varv(r)$ and $\dot{M}$ are largely shaped by the strength and stratification of $a_\mathrm{rad}$. 
In the subsonic region, the pressure term can be important as well. 
However, in the case of the studied cWR stars $a_\mathrm{rad}$ is significantly stronger than $a_\mathrm{press}$ \citep[see also][]{graefener2005,graefener2013}

In this work, we employ the PoWR$^\textsc{hd}$ branch to produce atmosphere models that resemble the spectral appearance of six Galactic WN4 stars while also reproducing the spectral energy distribution derived from photometry and flux-calibrated UV spectra. 
Besides the direct implications of the updated treatment of the $\varv(r)$, we also include significantly more elements than in most earlier studies of WN4b stars to ensure a realistic calculation of the radiative force.
An overview of these elements and their abundances can be found in Tab. \ref{tab:abundances}.
The additional constraints between the stellar and wind parameters due to the hydrodynamic coupling considerably reduce the number of free parameters and thus also the room for fine-tuning the spectral appearance. 
For example, any adjustment of the clumping factor has direct implications for the terminal velocity and the mass-loss rate. 
As a consequence, the width of an emission line cannot be changed without also affecting the electron scattering wings. 
While this can limit the visual ``fit'' between observation and model, the higher physical consistency nonetheless marks a significant added value for the interpretation of the resulting best-matching model.

The observed UV, optical, and IR spectra we employ in this work were also used in previous analyses \citep{hamann2006, sander2012}. All UV spectra were taken by the International Ultraviolet Explorer (IUE). The optical and IR spectra originate from the European Southern Observatory (ESO), La Silla (Chile), from the ``Deutsch-Spanisches Astronomisches Zentrum
(DSAZ)'' in Calar Alto (Spain), as well as from the Anglo-Australian Telescope (AAT) at Siding Spring (Australia). 
More detailed information for the optical and IR spectra of the six targets can also be found in \citet{hamann1995}. 
The optical narrow-band photometry data is from \citet{lundstrom1984}. In addition, we employ the photometry from Gaia DR3 and (in the case of WR6) PAN-STARRS \citep{chambers2016}. The IR broad-band photometry for our targets is taken from the 2MASS catalog \citep{skrutskie2006}.

Five of the six targets studied in this work are not showing any signs of a luminous companion and thus have been analysed as single objects. 
Three of our sample stars are known for intrinsic variability, most notably WR\,6 (aka HD\,50896 or EZ\,CMa), which is sometimes interpreted as a possible companion signature \citep[e.g.,][]{dsilva2022}. 
However, these variations are commonly interpreted as co-rotating interacting regions (CIRs), which were studied for several of our sample stars, namely WR\,1 \citep{stlouis2009}, WR\,6 \citep{stlouis2018} and WR\,58 \citep{chenee2011}.
Additionally, three of our objects have a nebula associated with them: WR\,6, WR\,7 and WR\,18 with respectively Sh2-308, NGC 2359, and NGC 3199 as emission nebulae containing hot gas \citep{toala2017,Camilloni+2024}. 
WR\,18 has recently been investigated with VLTI/GRAVITY by \citet{Deshmukh+2024}, finding a diffuse component contributing to the K-band but no indication of binarity.
We aimed for a spectral agreement of similar or improved quality over the previous analyses of our targets in \citet[hereafter \citetalias{hamann2019}]{hamann2019} and \citet[hereafter \citetalias{sander2019}]{sander2019}.
As the spectra of all six targets are wind dominated, we used the option introduced in \citet{Sander+2020} to fix the mass-loss rate and luminosity, while iterating over the stellar mass in the hydrodynamic iterations. 
Given the interdependencies and the long calculation times, a high-dimensional parameter grid is computationally extremely expensive. 
Hence, the iteration in the parameter space was performed under manual supervision with the best-fit model chosen by visual inspection, similar to prior studies performed for the targets.
The full visual comparisons can be seen in Appendix \href{https://zenodo.org/records/18220388}{B} for all six targets.

\section{Results and discussion}\label{sec:results}

Using hydrodynamically-consistent PoWR$^\textsc{hd}$ models (hereafter simply ``PoWR$^\textsc{hd}$ models''), we were able to successfully replicate observed spectra for our six targets.
For each target, the relevant parameters of the best-matching atmosphere models are shown in Table \ref{tab:stellar_params}. 
The parameters of the previous analysis using a grid approach and $\beta=1$ can be found in Table 1 from \citetalias{hamann2019} for WR\,1, WR\,6, WR\,7, WR\,18, and WR\,37 and in Table 1 from \citetalias{sander2019} for WR\,58.
The uncertainty values displayed in the figures of this study are based on the distance modulus (D.M.) uncertainties from \citet{crowther2023}, propagated via $\varepsilon\log_{10}\,L_\ast = 0.4\cdot\,\varepsilon\,\mathrm{D.M.}$, $\varepsilon\log_{10}\,R_\ast = 0.2\cdot\,\varepsilon\,\mathrm{D.M.}$, and $\varepsilon\log_{10}\,\dot{M} = 0.3\cdot\,\varepsilon\,\mathrm{D.M.}$. 
These latter three relations stem from the definition of the distance modulus $\mathrm{D.M.} = 5\log_{10}(d/10\,\mathrm{pc})$ (with $d$ the distance to the star), the Stefan-Boltzmann law ($L_\ast = 4\pi\,R_\ast^2\,\sigma_\mathrm{SB}\,\,T_\ast^4$) and the transformed radius (see Eq.\,\ref{eq:rtrans}).

\begin{table*}[h]
    \caption{Selected parameters from the best-fit PoWR$^\textsc{hd}$ models of the six targets in this study.}
    \label{tab:stellar_params}
    \centering
    \renewcommand{\arraystretch}{1.1}
    \begin{tabular}{l|cccccc}
        \hline\hline                       
        Parameter$^(a)$ & WR 1 & WR 6 & WR 7 & WR 18 & WR 37 & WR 58\\
        \hline
        $R_\ast\,[R_\odot]$ & $1.076\pm 0.04$ & $0.730_{-0.04}^{+0.06}$ & $0.564\pm 0.08$ & $1.207_{-0.06}^{+0.08}$ & $0.973\pm 0.08$ & $0.510_{-0.08}^{+0.09}$ \rule[1.05em]{0em}{0em}\\
        $R_\mathrm{crit}\,[R_\odot]$ & 1.101 & 0.528 & 0.584 & 1.234 & 1.001 & 0.475 \\
        $R_{\tau=2/3}\,[R_\odot]$ & 3.421 & 1.987 & 2.128 & 5.377 & 3.880 & 2.034 \\
        $T_\ast^(b)\, [\text{kK}]$ & $140\pm 2.4$ & $143\pm 1.6$ & $145\pm 2.7$ & $140\pm 1.6$ & $143\pm 0.8$ & $144\pm 2.6$ \\
        $T_\mathrm{eff} (R_\mathrm{crit})\,[\text{kK}]$ & 138.4 & 141.0 & 142.5 & 138.4 & 141.0 & 141.4 \\
        $T_{\tau=2/3}\,[\text{kK}]$ & 78.5 & 86.7 & 74.6 & 66.3 & 71.6 & 72.1 \\
        $\log_{10}(L_\ast\,[L_\odot])$ & $5.6\pm 0.08$ & $5.3_{-0.08}^{+0.12}$ & $5.1\pm 0.16$ & $5.7_{-0.12}^{+0.16}$ & $5.6\pm 0.16$ & $5.0_{-0.16}^{+0.18}$ \\
        $M_\ast\,[M_\odot]$ & $19.6\pm 1.25$ & $10.3 \pm 0.27$ & $6.0\pm 0.56$ & $22.3\pm 3.39$ & $14.5\pm 1.28$ & $4.9\pm 0.84$ \\
        $\log_{10}\left(\dot{M}\,\left[M_\odot\,\mathrm{yr}^{-1}\right]\right)$ & $-4.9\pm 0.06$ & $-5.1_{-0.06}^{+0.09}$ & $-5.2\pm 0.12$ & $-4.7_{-0.09}^{+0.12}$ & $-4.8\pm 0.12$ & $-5.3_{-0.12}^{+0.14}$ \\
        $\log_{10}\left(\dot{M}_\mathrm{t}\,\left[M_\odot\,\mathrm{yr}^{-1}\right]\right)$ & $-3.9\pm 0.12$ & $-4.0_{-0.12}^{+0.18}$ & $-4.0\pm 0.24$ & $-4.1_{-0.18}^{+0.24}$ & $-4.0\pm 0.24$ & $-4.0_{-0.24}^{+0.28}$ \\
        $\varv_\infty\,\left[\mathrm{km}\,\mathrm{s}^{-1}\right]$ & $2023\pm 288$ & $1759\pm 179$ & $1328\pm 175$ & $1603\pm 216$ & $1788\pm 117$ & $1225\pm 300$ \\
        $D_\infty^(c)$ & 70 & 50 & 20 & 15 & 20 & 20\\
        \hline
    \end{tabular}
        \tablefoot{
        \tablefoottext{a}{All models have a microturbulent velocity of $\varv_\mathrm{mic} = 30$ km s$^{-1}$.}
        \tablefoottext{b}{The effective temperature at the inner boundary $R_\ast$, set at $\tau_\text{Ross,cont}=20$.}
        \tablefoottext{c}{We adopt a depth-dependent clumping law \citep{hillier2003} where we use 100 km s$^{-1}$ as the onset velocity $\varv_\mathrm{cl}$ (250 km s$^{-1}$ for WR\,58) and where $D_\infty$ represents the clumping when $\varv >> \varv_\mathrm{cl}$.}
    }
\end{table*}

\subsection{Stellar parameter revision}\label{subsec:param_differences}

\begin{figure}[h]
	\centering
    \includegraphics[width=\hsize]{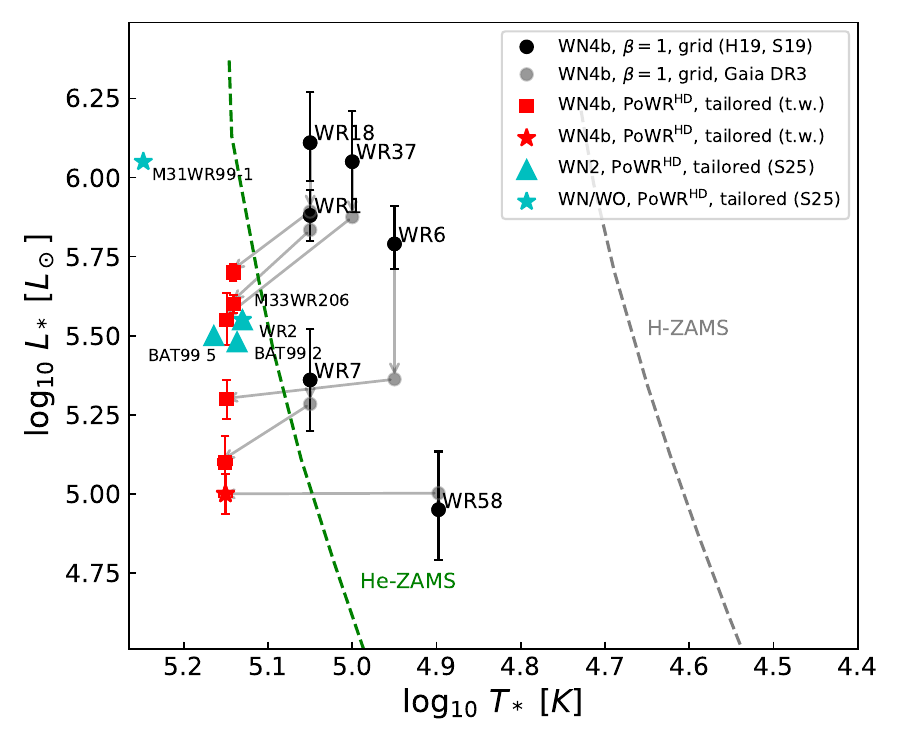}
	\caption{HR-diagram of the WN4b sample with their parameters from \citetalias{hamann2019} and \citetalias{sander2019} (black circles), updated parameters using only corrected parallaxes from Gaia DR3 (grey circles), and parameters from modelling with PoWR$^\textsc{hd}$ in this work (red). 
    Gaia DR3 corrections are depicted by arrows from black circles to grey circles. 
    Differences in addition to the distance corrections are illustrated with grey arrows from the grey circles to the red data points.
    For comparison, we also show the stars from \citetalias{sander2025} (cyan) as well as the H-ZAMS from \citet{ekstroem2012} and the He-ZAMS from \citet{langer1989} as the grey and green dashed lines, respectively.}
	\label{fig:hrd}
\end{figure}

There are substantial differences between the best-matching PoWR$^\textsc{hd}$ models and the previous analyses for all six targets. 
Beside the fact that the older studies are based on assigning mainly grid models using $\beta=1$ and containing significantly fewer elements than the current PoWR$^\textsc{hd}$ models, there is also updated parallax information due to the release of Gaia DR3. 
In this work, we use the new distances given in \citet{crowther2023} and the corresponding online WR catalogue\footnote{\url{http://pacrowther.staff.shef.ac.uk/WRcat/}} derived from the Gaia DR3 parallaxes including local zero point corrections by \citet{Lindegren+2021} and \citet{MaizApellaniz2022}. 
This results in considerable luminosity updates for some of the targets, mostly downward as the previous parallaxes with higher uncertainties typically led to an overestimation of the distance.
Together with the switch to PoWR$^\textsc{hd}$ models, significant position shifts in the HR-diagram compared to \citetalias{hamann2019} and \citetalias{sander2019} are obtained as shown in Fig.\,\ref{fig:hrd}. 
The figure shows a two-step shift: (1) the changes in $L_\ast$ solely due to the updated Gaia parallaxes and (2) the changes in both $T_\ast(\tau=20)$ and $L_\ast$ due to the different modelling we employ in this study.
Here we note that, due to the updated distances, the $\dot{M}$ values of the \citetalias{hamann2019} and \citetalias{sander2019} models will also be affected (see e.g. Sect. 4 in \citetalias{hamann2019}).
As evident from Table\,\ref{tab:stellar_params}, the effective temperatures at the critical radius $T_\mathrm{crit}$ are very similar to $T_\ast$. 
This means that their winds must be launched rather deep as the radii corresponding to $\tau=20$ and the critical radius reflecting the wind onset are not very different.
Moreover, the corresponding effective temperatures are now all very similar to $\sim$$140$\,kK and beyond the He-ZAMS line \citep{langer1989}.
With the exception of WR\,58, the $L_\ast$-values on the other hand all decrease to various degrees when considering the new Gaia data. 
For comparison we also display the WN2 and WN/WO stars from \citet[hereafter \citetalias{sander2025}]{sander2025} which have recently been analysed with the same methodology. 
With the exception of M31WR\,99-1, all \citetalias{sander2025} targets coalesce around a similar $T_\ast$, albeit with a smaller spread in $L_\ast$.
We also note that both WR\,7 and WR\,58 have comparatively low luminosities for the bulk of WR stars and fall in the range of intermediate-mass stripped stars. 
The values we derive here are close to the (partially) stripped stars in \citet{gotberg2023} and \citet{ramachandran2024} in the SMC, however, with significantly higher mass-loss rates.

While the PoWR$^\textsc{hd}$ models line up with very similar $T_\ast$ values in Fig.\,\ref{fig:hrd}, a much wider range of temperatures was obtained in the grid analyses in \citetalias{hamann2019} and \citetalias{sander2019}. 
This was due to parameter degeneracies that limit constraining several stellar and wind parameters only to $\dot{M}_\text{t} \propto R_\mathrm{t}\,T^2$ (see also Eq.\,\eqref{eq:rtrans}) in the case of dense winds. 
This degeneracy can be broken with the PoWR$^\textsc{hd}$ modelling. 
As the wind structure is uniquely solved per atmosphere model, we can derive the effective temperature at the hydrostatic surface, for which we use the critical point as defined in \citet{sander2017} as a reference radius. 
Moreover, as a specific set of wind parameters is coupled to a set of stellar parameters, it is not possible to change the emerging spectrum without changing the underlying stellar parameters, which provides us an indirect estimate of the otherwise inaccessible stellar mass.
As hydrogen-free WN stars are close to the theoretical concept of a ``helium star'' \citep{langer1989}, our targets should be close to or slightly hotter than the He zero age main sequence (ZAMS). 
This is indeed the case when inspecting the $T_\mathrm{crit}$ values for our WN4b sample as well as most of the \citetalias{sander2025} sample. 
Confirming the suggestions from the early WC prototype model by \citet{graefener2005}, we can thus conclude that at least for WN2 and WN4b stars, hydrodynamically-consistent atmosphere modelling is able to resolve the ``WR radius problem''.

\begin{figure}[t]
	\centering
    \includegraphics[width=\hsize]{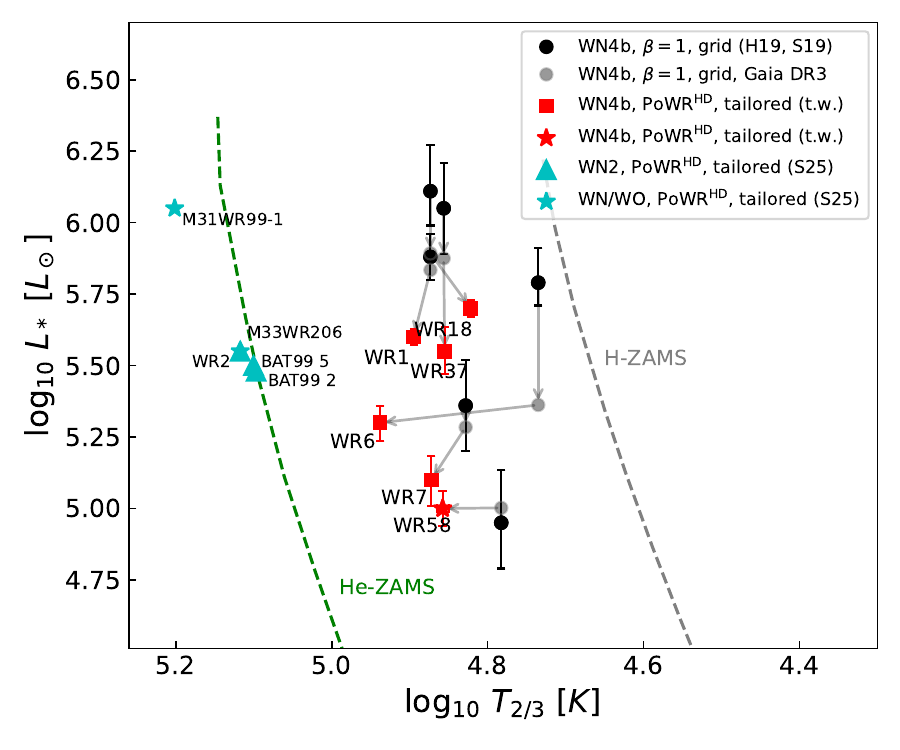}
	\caption{Same as in Fig. \ref{fig:hrd}, but now with the effective temperature $T_{2/3}$ corresponding to an optical depth of $\tau=2/3$, on the x-axis.}
	\label{fig:hrd_t23}
\end{figure}

Notably, when considering the photospheric temperatures $T_{2/3} = T_\text{eff}(\mathrm{\tau=2/3})$ instead of $T_\ast$, we obtain significantly smaller shifts in the HRD compared to the older studies as shown in Fig.\,\ref{fig:hrd_t23}, in particular along the temperature axis. 
This illustrates that for most of our targets the older models with $\beta = 1$ were reasonably sufficient in describing the layers from where the observed spectrum is emerging. 
We will discuss this further in Sect. \ref{subsec:wind_comparison} when directly comparing the wind structures. 
When comparing $T_\mathrm{crit}$ to $T_\mathrm{\tau=2/3}$ for the WN4b stars, i.e., the effective temperatures of the wind onset and of the emergent continuum, the difference is roughly a factor of two (cf.\ Table\, \ref{tab:stellar_params}). 
This reflects how extended the atmospheres of the WN4b stars are, in sharp contrast to the PoWR$^\textsc{hd}$ results for the WN2 and WN/WO stars in \citetalias{sander2025}, where the two temperatures are very similar. 
This is also evident from their position in the HRD (see Fig.\,\ref{fig:hrd_t23}), where the WN2 and WN/WO stars reside close to the He-ZAMS. 

In terms of chemical abundance determination, the six targets do not deviate much from the bulk of Galactic WN stars. 
Initially using the abundances from the PoWR WN-star model grids \citep{hamann2004grids,todt2015}, with some small variations in the C and N abundances in order to assure certain spectral lines are modelled well.
One such example is the case of the \ion{C}{iv} $\lambda\,5808\,\AA$ line for WR58, shown in Fig. \ref{fig:wr58_full}, where an increased C abundance is crucial as WR58 is a known WN/WC star \citep{vdH2001}. 
For the other elements included in the wind modelling, where we have used solar abundances from \citet{asplund2009}, we refer to Tab.\,\ref{tab:abundances} for the list of derived mass fractions and the modelled ions per element.

\subsection{Wind structure comparison}\label{subsec:wind_comparison}

\begin{figure}[h]
	\centering
    \includegraphics[width=\hsize]{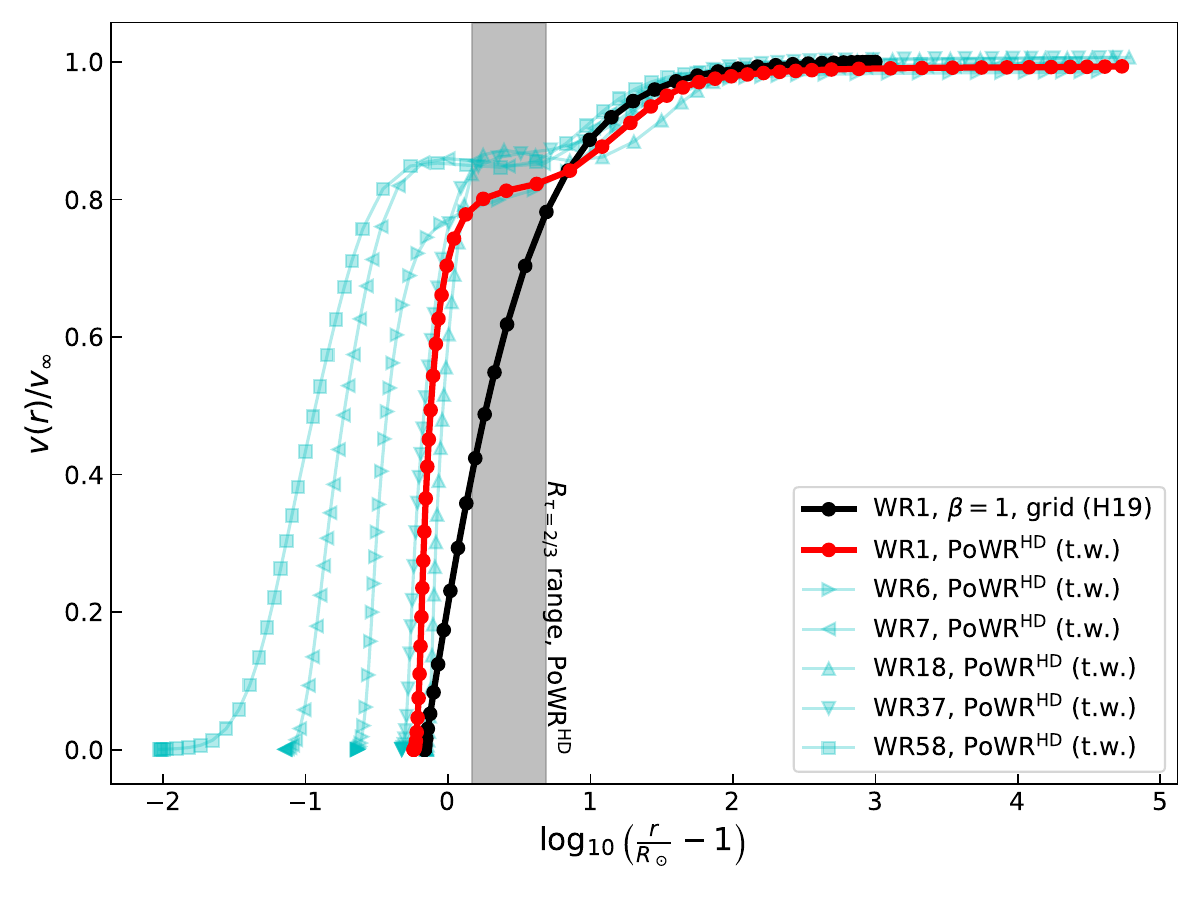}
	\caption{Velocity stratification $\varv(r)$ comparison of the best-fit WN4b $\beta=1$ grid model (\citetalias{hamann2019}, black connected circles) and the PoWR$^\textsc{hd}$ models in this work (red connected circles) of WR1.
    The $\varv(r)$ of the remaining 5 targets in this work are shown in cyan for comparison.
    The range of photospheric radii of the PoWR$^\textsc{hd}$ models at optical depth $\tau=2/3$ $R_{2/3}$ is shaded in grey.
    The difference in outermost radii between the grid models and the PoWR$^\textsc{hd}$ models is due to the former being traditionally modelled up to 1\,000 $R_\odot$, with the latter up to 50\,000 $R_\ast$.
    } 
	\label{fig:velo_comp}
\end{figure}

In Fig.\,\ref{fig:velo_comp}, we take a closer look at the velocity stratification for our six targets. 
In order to better compare the slopes, we scale the velocities to the respective $\varv_\infty$ of the corresponding model. 
For comparison, we also show the old model for WR\,1 from \citet{hamann2006} and \citetalias{hamann2019} using $\beta = 1$.
In the outer wind, past $10$ to $50\,R_\odot$, the slopes of the velocity laws essentially align with the $\beta$-law.
However, in the inner wind, there is a significant deviation from a simple $\beta$-law as the velocity solutions from the PoWR$^\textsc{hd}$ models show a much steeper increase followed by a plateau or even a small decrease of the wind speed around $\varv \approx 0.85\,\varv_\infty$. 
Our obtained $\varv(r)$-stratifications generally align with the type of shape obtained for the fiducial early-type WN model presented in \citet{Sander+2020}. 
The ability to also yield non-monotonic solutions was later added in \citet{sander2023temp} and is necessary for some of the targets here, e.g., for WR\,18 and WR\,58.
Note that the different total radial extension is only due to the different outer boundary choices in the modelling. 
While the traditional grid models are calculated up to 1\,000\,$R_\odot$, the PoWR$^\textsc{hd}$ models were calculated up to $50\,000\,R_\ast$.
This difference does, however, not significantly impact the resulting optical or UV spectra. 
Interestingly, the photospheric radii $R_{2/3}$, i.e., the radii at optical depth $\tau=2/3$,  coincide with the velocity stratification plateaus in the PoWR$^\textsc{hd}$ models. 
To first order, the remaining velocity field outwards of $R_{2/3}$ can be described by a $\beta$-law, explaining the former success in the overall reproduction of the spectral features using $\beta$-law models. 
Moreover, it implies that there are still significant changes in the optically thin regime of the wind before $\varv_\infty$ is reached.

\begin{figure}[h]
	\centering
    \includegraphics[width=\hsize]{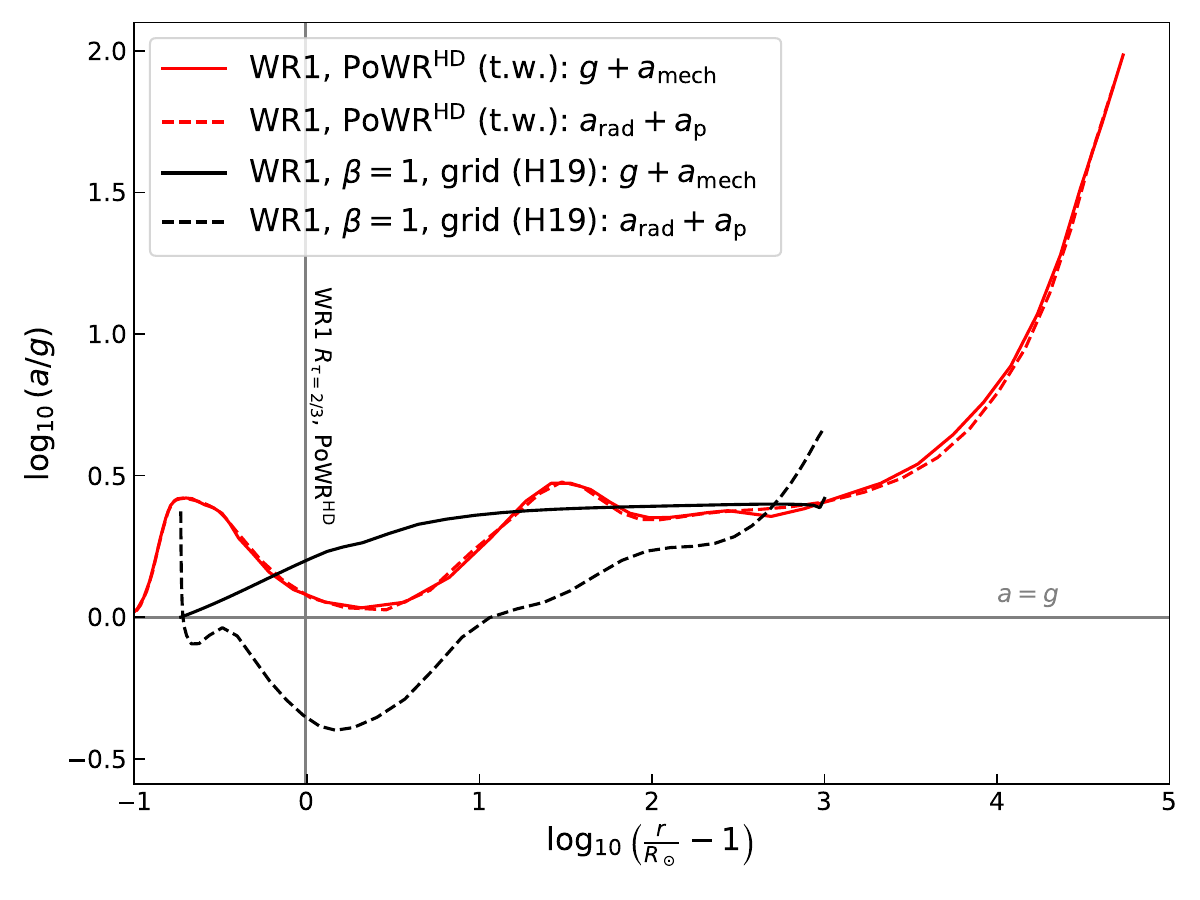}
	\caption{Acceleration stratification comparison between the WN4b $\beta=1$ grid model (\citetalias{hamann2019}, black lines) and the PoWR$^\textsc{hd}$ model (red lines) of WR1. 
    The main wind deceleration terms (gravity $g$ and wind inertia $a_\mathrm{mech}$) are solid lines, while the main acceleration terms (radiative force $a_\mathrm{rad}$ and pressure gradient $a_\mathrm{press}$) are dashed lines.
    The grey horizontal line depicts the accelerations equal to gravity.
    The reason for the difference in outermost radii is the same as in Fig. \ref{fig:velo_comp}.}
	\label{fig:acc_comp}
\end{figure}

To understand the origin of the major differences in the inner wind, it is helpful to visualize the force balance for an exemplary hydro model and compare it to the situation in a $\beta$-type model.
In Fig.\,\ref{fig:acc_comp}, we depict the main acceleration and deceleration terms: gravity $g$ and inertia $a_\mathrm{mech}$ slow down the wind material, while the pressure gradient $a_\mathrm{p}$ and radiative acceleration $a_\mathrm{rad}$ push the wind outwards. The figure shows the total deceleration ($g + a_\mathrm{mech}$) and the total acceleration ($a_\mathrm{p} + a_\mathrm{rad})$ align for the PoWR$^\textsc{hd}$ model as hydrodynamic equilibrium is enforced. 
In contrast, the prescribed $\beta$-velocity law used in \citetalias{hamann2019} and \citetalias{sander2019} do not account for the radiative acceleration in the inner wind due to the iron M-shell opacities.
The location of $R_{2/3}$ is marked in Fig.\,\ref{fig:acc_comp} as well, illustrating again that the spectra are formed outwards of this rapid inner acceleration and explaining why $\beta$-laws are yielding a reasonable spectrum despite the shortcomings in inferring the correct wind onset radii.

\subsection{Mass-loss rates}\label{subsec:mdot}

\begin{figure}[ht]
	\centering
    \includegraphics[width=\hsize]{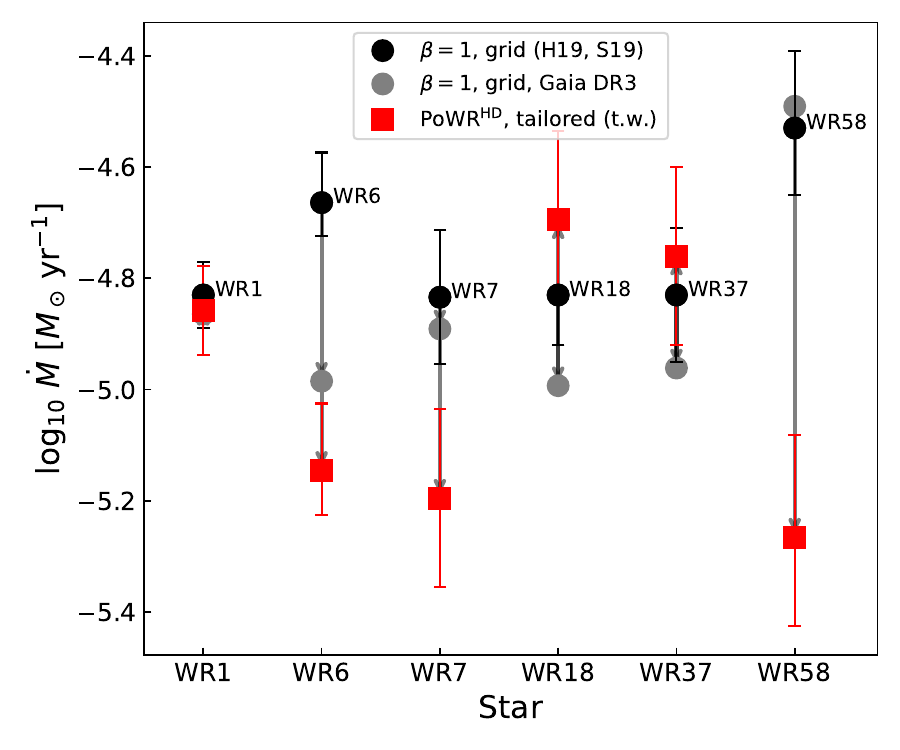}
	\caption{Mass-loss rate $\dot{M}$ comparison between the WN4b $\beta=1$ grid models (\citetalias{hamann2019} \& \citetalias{sander2019}, black circles) and the PoWR$^\textsc{hd}$ models in this work (red squares). 
    In similar fashion to Figs. \ref{fig:hrd} and \ref{fig:hrd_t23}, the Gaia DR3 distance correction is applied on the grid models (grey circles) separate from the $\dot{M}$ updates in this work (see Subsect. \ref{subsec:mdot}). 
    The two-step parameter difference is denoted with the grey arrows per target.}
	\label{fig:mdot_comp}
\end{figure}

The mass-loss rates $\dot{M}$ from the grid models in \citetalias{hamann2019} and \citetalias{sander2019}, and the PoWR$^\textsc{hd}$ models are shown in Fig. \ref{fig:mdot_comp}.
In order to identify the influence of the different atmosphere modelling, we need to separate again the correction resulting from the Gaia DR3 parallaxes applied to the older grid model solution. 
The classic WR model grids were calculated in two-dimensional planes spanning $T_\ast$ and the transformed radius:

\begin{equation}\label{eq:rtrans}
    R_\mathrm{t} = R_\ast \cdot\left(\frac{\varv_\infty}{2500\,\mathrm{km}\mathrm{s}^{-1}} \Bigg/ \frac{\dot{M}\sqrt{D}}{10^{-4}\,M_\odot\,\mathrm{yr}^{-1}}\right)^{2/3},
\end{equation}

\noindent where $D$ is the clumping factor. 
The definition of $R_\mathrm{t}$ stems from the early insight that models with varying $\dot{M}$, $R_\ast$ and $\varv_\infty$ but the same $R_\mathrm{t}$ have very similar emission line equivalent widths \citep{schmutz1989}. 
Due to the dependence on $R_\ast$ (and hence on $L_\ast$), the mass-loss rates need to be scaled as well in order to keep $R_\mathrm{t}$ and thus the normalized spectrum when luminosities change due to updated parallaxes.
As evident from Fig.\,\ref{fig:mdot_comp}, there is no systematic shift in $\dot{M}$ when shifting from the grid models to PoWR$^\textsc{hd}$ models. While WR1 shows hardly any changes at all, WR6, WR7, and WR58 show a significant decrease in $\dot{M}$ for our models. In some cases (e.g., WR18 and WR37, which have their $\dot{M}$ values increased compared to \citetalias{hamann2019}), the effect from the different modelling is contrary to the shift resulting from the updated distance.

\begin{figure}[htb]
	\centering
    \includegraphics[width=\hsize]{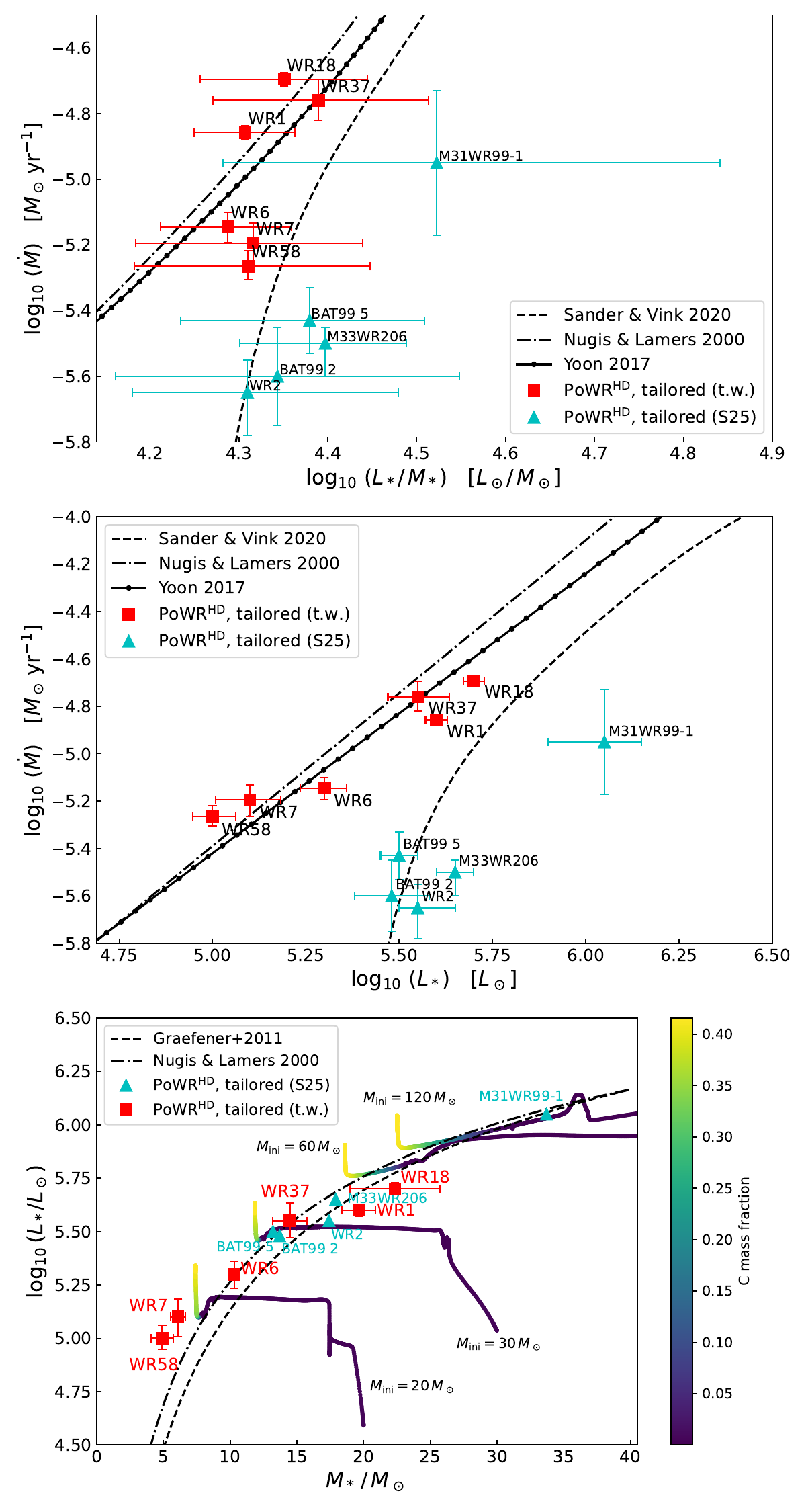}
	\caption{The top panel shows $\dot{M}$ values of the PoWR$^\textsc{hd}$ WN4b models in this work (red squares) and the \citetalias{sander2025} targets (cyan triangles) in terms of $L_\ast / M_\ast$. 
    For comparison, the $\dot{M}$ laws from \citet{nugis_lamers2000}, \citet{yoon2017} and \citet{SanderVink2020} are displayed. 
    The middle panel shows the same $\dot{M}$-values, now in terms of $L_\ast$ and compared with the same predictions as in the top panel.
    In the bottom panel, the model $L_\ast$ values are shown in terms of $M_\ast$. 
    The $L_\ast \sim M_\ast$ relations from \citet{nugis_lamers2000} and \citet{graefener2011} are shown as well, along with the evolutionary tracks (with rotation) from \citet{chieffi2013} for initial masses 20, 30, 60 and 120 $M_\odot$.
    The evolutionary tracks are colour-coded by carbon surface abundances in mass fractions.}
	\label{fig:mdotlaw_comp}
\end{figure}

With the updated mass-loss rates and the stellar parameters determined, we can test whether these combinations are predicted by commonly applied mass-loss recipes.
In Fig.\,\ref{fig:mdotlaw_comp}, we compare the descriptions by \citet{nugis_lamers2000}, the combined recipe by \citet[using $f_\text{WR} = 1.0$]{yoon2017}, and the more recent formula from \citet{SanderVink2020} to the values we determined for our six targets as well as to the stars from \citetalias{sander2025}. 
With the exception of M31WR\,99-1, all of the targets have $T_\text{crit} \approx 141\,$kK and thus we do not consider the temperature correction from \citet{sander2023temp} when plotting the recipe curve (which would anyhow be target-dependent).
In the uppermost panel of Fig.\,\ref{fig:mdotlaw_comp}, we plot $\dot{M}$ versus $L_\ast / M_\ast$. 
The derived WN4b star parameters show quite some spread in this plane with larger uncertainties along the x-axis due to both $L_\ast$ and $M_\ast$ having inherent uncertainties. 
Notably, all stars are located above the line from \citet{SanderVink2020} for solar metallicity. 
This is particularly interesting as the basic methodology to derive the recipe was the same, namely dynamically-consistent modelling with PoWR$^\textsc{hd}$. 
While some of the targets reach the curve within the uncertainties, others clearly do not. 
\citet{SanderVink2020} assumed the mass-luminosity-relation from \citet{graefener2011} for their targets, which might not be the case for some of the WN4b stars. 
We will examine this further below. 
Interestingly, the WN2 star WR\,2 is precisely on the \citet{SanderVink2020} curve for solar metallicity. 
The other hotter WN and WN/WO stars are rightward of the curve, which is to be expected for subsolar metallicity. 
The WN/WO star M31WR\,99-1, as briefly mentioned above, is significantly off but expected to have a weaker mass-loss rate due to its considerably hotter temperature.

The mass-loss recipes from \citet{nugis_lamers2000} and \citet{yoon2017} do not have an $L/M$-dependence, but only an $L$-dependence. 
To still compare them in the $\dot{M}$-$L/M$-plane, we use Eq.\,(5) from \citet{nugis_lamers2000} to calculate the masses they would have assigned to those luminosities. 
Beside altering the offset, their formula is taken from \citet{schaerer1992} and thus based on stellar structure calculations. 
We apply the same relation also to the \citet{yoon2017} recipe in order to be able to compare their performance relative to each other.
Interestingly, one half of the WN4b sample matches rather well with the such calibrated \citet{nugis_lamers2000} relation, while the other stays above a factor of two in $\dot{M}$ below the curve. 
The second half of the sample is even overpredicted by the \citet{yoon2017} relation, although -- for this limited sample -- one could conclude that it provides a reasonable compromise. 
However, the relations fail for the sample of WN2 and WN/WO stars from \citetalias{sander2025} which have much lower mass-loss rates. 
Considering the BAT99 targets to have approximately $0.5\,Z_\odot$, \citet{yoon2017} would predict a shift of the curve by about $-0.18\,$dex in $\dot{M}$ due to $Z^{0.6}$. 
The $Z^{0.85}$ prediction from \citet{VinkdeKoter2005} would yield $-0.25\,$dex. 
Instead, downward shifts of about $0.7$ to $0.8\,$dex would need to be applied to reach the derived locations in the $\dot{M}$-$L/M$-plane. 
We can thus conclude that the empirically calibrated mass-loss recipes from both \citet[with $f_\text{WR} = 1.0$]{yoon2017} and \citet{nugis_lamers2000} give reasonable results if the targets are similar to the bulk of the sample that was used to derive them. 
However, they struggle to predict the correct $\dot{M}$ for objects that are considerably different and comparatively rare. 
The theoretical/model-derived description can include such rare cases, but in turn struggles when objects are different from their basic assumptions.

Given that the recipes from \citet{yoon2017} and \citet{nugis_lamers2000} explicitly only include an $L$-dependence and are also employed that way in many evolution calculations, we also compare our results in the $\dot{M}$-$L$-plane (cf.\ middle panel of Fig.\,\ref{fig:mdotlaw_comp}). 
In order to have the pure $L_\ast$ dependence for the \citet{SanderVink2020} recipe, we employ Eq.\,(13) in \citet{graefener2011} to calculate stellar masses. 
For plotting the recipes, we assumed $Y = 0.98$ and $Z = 0.014$, in line with our empirical abundance determinations.
Both \citet{nugis_lamers2000} and \citet{yoon2017} do an excellent job in mapping the derived results although it seems that the empirical points hint at a less steep slope.  
Considering again the uppermost panel of Fig.\,\ref{fig:mdotlaw_comp} and the error margins, it is clear that all WN4b targets (and even WR\,2) have very similar $L/M$ ratios. 
As the stars have nearly the same $T_\text{crit}$, one might say that an $L$-dependence is sufficient. 
However, the proximity to the Eddington Limit also needs to be accounted for, and thus an $L$-dependence has to be in addition to an $L/M$-dependence rather than instead.

In the bottom panel of Fig.\,\ref{fig:mdotlaw_comp}, we now study the mass-luminosity plane, where we plot both the \citet{graefener2011} relation for hydrogen-free stars \citep[inherent to the predictions from][]{SanderVink2020} and the \cite{nugis_lamers2000} relation drawn from \citet{schaerer1992}. 
While the WN4b stars are close to the two curves, none of them is actually on the He-ZAMS relation from \citet{graefener2011}. 
The lower mass stars in our sample (WR58 and WR7) show lower masses than expected for a star on the He-ZAMS, which could indicate that these stars are a bit further evolved than the rest of the sample. 
This effect can also be seen in the evolutionary tracks from \citet{chieffi2013}, which we plot for comparison and annotate with their respective initial masses and surface carbon mass fractions. 
Towards the end of core-He burning, the stars are predicted to get more luminous again. 
Yet, as we will see in Sect.\,\ref{subsec:evolution}, this is usually obtained together with an increase in the temperature, which we do not observe. Nonetheless, the evolved interpretation would be in line by the fact that WR\,7 and WR\,58 are located above the $\dot{M}(L/M)$ prediction for He-ZAMS-based models (middle panel of Fig.\,\ref{fig:mdotlaw_comp}), while being ``normal'' with respect to $\dot{M}_\text{t}(L/M)$, which we will discuss further below.

While lower masses for a given luminosity are expected from more evolved objects, the locations of WR\,1 and WR\,18 in the lower panel of Fig.\,\ref{fig:mdotlaw_comp} are indicating the opposite, namely an excess in mass compared to their luminosity, albeit with some uncertainty. 
As both stars have clearly no considerable amount of remaining hydrogen, the HeZAMS mass should be an upper limit for a given luminosity. 
One explanation could be an overestimation of the mass from the PoWR$^\textsc{hd}$ models, which \citet{gonzalez-tora+2025} concluded when comparing PoWR$^\textsc{hd}$ models with 3D radiation-hydrodynamic models from \citet{moens2022}. 
In their study, this effect was on the order of 10\% in the mass, which could be just enough to bring the targets back onto the \citet{graefener2011} curve. 
However, this would in turn also increase the luminosity excess for WR\,58 and WR\,7. 
Alternatively, our six targets could indicate that the $L$-$M$ relation for hydrogen-free is actually a bit different from current structural predictions, but the small sample and the systematic uncertainties prevent us from drawing bigger conclusions here.

\begin{figure}[h]
	\centering
    \includegraphics[width=\hsize]{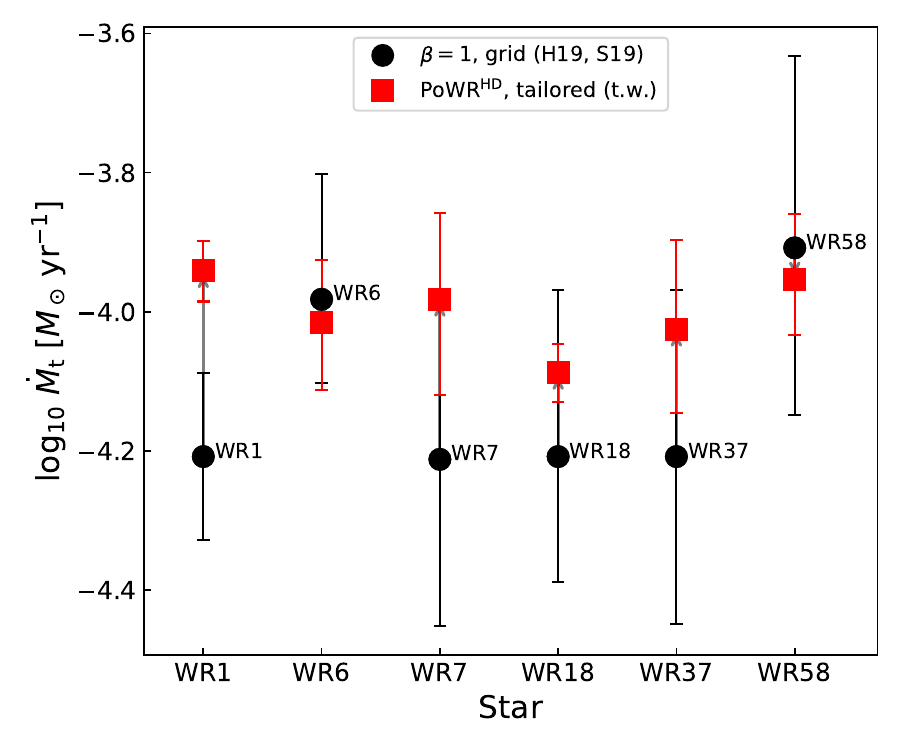}
	\caption{The transformed mass-loss rates ($\dot{M}_\mathrm{t}$, see eq. \eqref{eq:mdottrans}) of the WN4b $\beta=1$ grid models (\citetalias{hamann2019} \& \citetalias{sander2019}, black circles) compared with the PoWR$^\textsc{hd}$ models in this work (red squares). 
    The grey arrows denote the parameter difference between the two model sets.
    Here, no distance correction as in e.g. Fig. \ref{fig:mdot_comp} is required.}
	\label{fig:mdottrans_comp}
\end{figure}

\begin{figure}[h]
	\centering
    \includegraphics[width=\hsize]{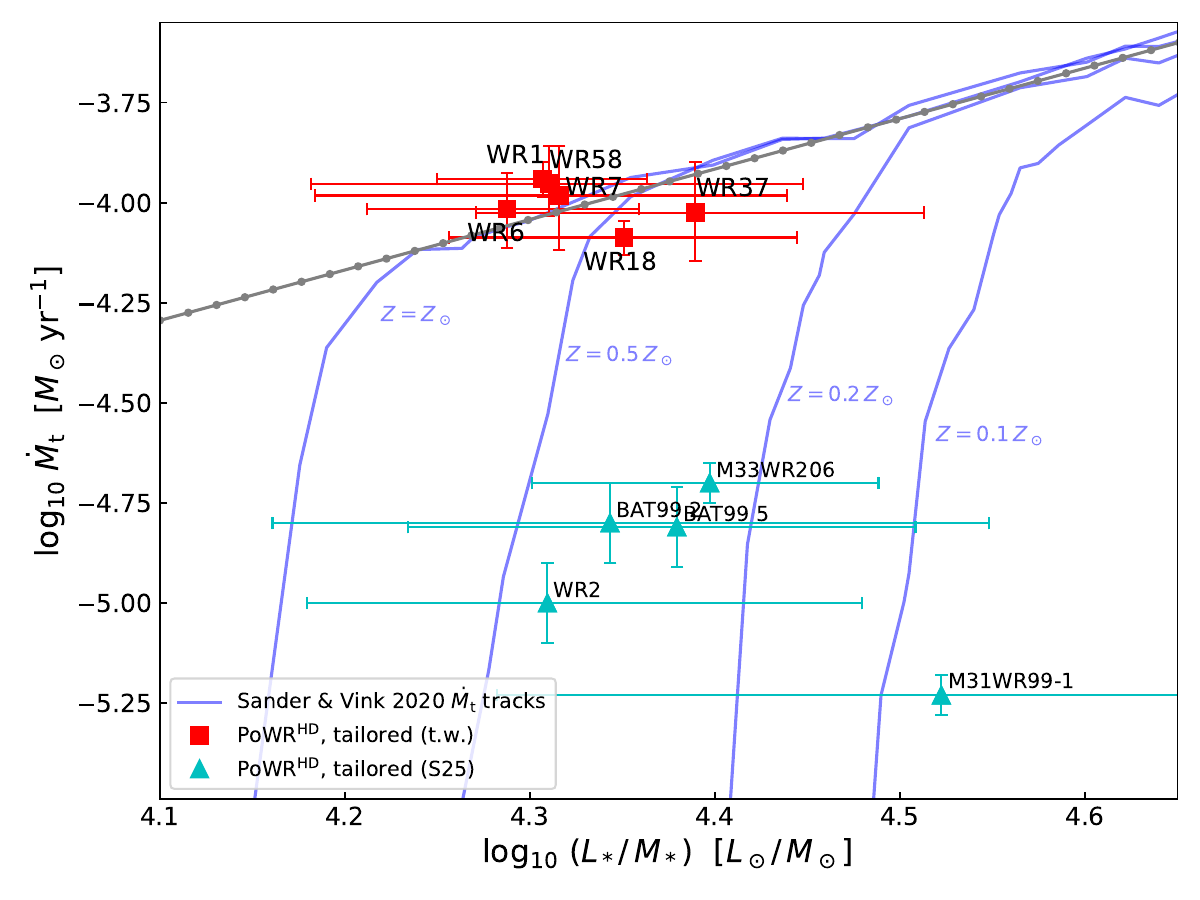}
	\caption{Comparison of the WN4b PoWR$^\textsc{hd}$ models in this work (red squares), the \citetalias{sander2025} models (cyan triangles) in $\dot{M}_\mathrm{t}$ - $L_\ast / M_\ast$ space.  
    In blue are the $\dot{M}_\mathrm{t}$ tracks from \citet{SanderVink2020} depicted for $Z = 0.1$, 0.2, 0.5 $Z_\odot$ and $Z = Z_\odot$. 
    The thick gray dotted line indicates the fit in the dense-wind regime derived in \citet{SanderVink2020}.}
	\label{fig:mdottrans_lm_comp}
\end{figure}

Finally, we compare the transformed mass-loss rate \citep[defined in][]{graefener2013}:
\begin{equation}\label{eq:mdottrans}
    \dot{M}_\mathrm{t} := \dot{M}\,\sqrt{D} \cdot \left(\frac{1000\,\mathrm{km}\,\mathrm{s}^{-1}}{\varv_\infty}\right)\left(\frac{10^6\,L_\odot}{L_\ast}\right)^{3/4},
\end{equation}
which essentially represents the mass-loss if the star had $L_\ast=10^6\,L_\odot$, $\varv_\infty=1000\,\mathrm{km}\,\mathrm{s}^{-1}$ and $D=1$. 
This quantity is a more elegant alternative to $R_\text{t}$ and enables comparisons of objects with different luminosities. 
As this quantity does not change with distance updates when using grid models, the changes visualized in Fig.\,\ref{fig:mdottrans_comp}  reflect the necessary adjustments when transitioning from $\beta$-type models to PoWR$^\textsc{hd}$ models. 
Here, one needs to be aware that PoWR$^\textsc{hd}$ does no longer enable manual adjustments of $\varv_\infty$ as this is an outcome of the basic stellar parameters and the wind solution. 
Therefore, the best-fit model is often a compromise where $\varv_\infty$ is not perfectly tailored to the widths of emissions lines and P\,Cygni absorptions. 
Already from this effect, some variation in $\dot{M}_\text{t}$ compared to models where $\varv_\infty$ is a free parameter is to be expected.

\citet{SanderVink2020} found a linear trend of $\dot{M}_\mathrm{t}$ versus $L/M$ in the regime of optically thick winds with a $Z$-dependent breakdown when the winds eventually become optically thin. 
In Fig.\,\ref{fig:mdottrans_lm_comp}, we show the derived positions of our sample stars compared to both the linear relation as well as tracks for specific metallicities arising from individual sets of models. 
Notably, WN4b stars align very well with the $\dot{M}_\mathrm{t}$ prediction. WR\,7 and WR\,18 are shifted a bit downwards, which could indicate some issues with the derived parameters, but are still clearly in the dense-wind regime. 
In contrast, the WR sample from \citetalias{sander2025} shows optically thin winds and the locations in this diagram indicate that these stars might indeed have metallicities between LMC and SMC level -- again with the exception of M31WR\,99-1 which is clearly off, but also has a much higher $T_\text{crit}$.

\subsection{ionising photon rates}

As our WN4b stars have high effective temperatures and large luminosities, they are expected to be strong ionising sources. 
As illustrated in Fig.\,\ref{fig:qhe_comparison}, all of our targets are indeed strong sources of ionising photons beyond the hydrogen and \ion{He}{i} ionization edge. 
When comparing with the older analysis results, we see that the obtained changes are all in the direction of the adjustments in the luminosities of the targets due to the distance updates discussed above. A significant additional shift from the change to the PoWR$^\textsc{hd}$ models is only visible for targets that also had a further luminosity shift in the HRD (i.e., WR\,1, WR\,7, and WR\,37, cf.\ Fig.\,\ref{fig:hrd}).
This outlines that the difference in the velocity field and in particular the inner wind structure hardly affect these integrated quantities, implying that the common usage of ionising photon calculations from $\beta$-type atmosphere models in population synthesis is not a problem as such. 
However, our results also underline that the ionising photon production is not directly related to the hydrostatic radii of the stars.
The expanding photospheres need to be taken into account, which is an open problem in current population synthesis modelling as the assignment of atmosphere models to predictions from evolutionary tracks is a highly non-trivial problem \citep[see, e.g.,][]{groh2014,josiek2025,roy2025}.

When inspecting the number of ionising photons beyond the \ion{He}{ii} edge, we obtain values on the order of only $10^{41}\,\mathrm{s}^{-1}$, many orders of magnitude below what we obtain for $Q_\mathrm{H}$ and $Q_\mathrm{\ion{He}{I}}$. 
This is not unexpected as the winds of all of our targets are so dense that these photons cannot escape. 
This is in stark contrast to the WN2 and WN/WO stars analysed in \citetalias{sander2025}, where the $Q_\mathrm{\ion{He}{II}}$ values are comparable to the $Q_\mathrm{\ion{He}{I}}$ values of the WN4s here. 
In the model sequences calculated in \citet{sander2023temp}, a characterised transformed mass-loss rate of $\dot{M}_\text{t} \approx 10^{-4.5}\,M_\odot\,\mathrm{yr}^{-1}$ was found for the transition between WR stars that are opaque and transparent to \ion{He}{ii} ionising photons \citep[see also][]{gonzalez-tora2025_lvm}. 
Indeed, our WN4b stars have $\dot{M}_\text{t}$ values considerably above this value while the WN2 and WN/WO stars from \citetalias{sander2025} show values below this limit. 
At least for the studied regime of WRs on or close to the He-ZAMS we can thus confirm the suitability of this value as a transition diagnostic.

\subsection{Comparison with evolutionary tracks}\label{subsec:evolution}

While a detailed discussion about the origin of the WN4 stars easily warrants its own study, our derived HRD positions allow us to compare them to typical sets of published evolution tracks. 
Typically, evolution models add a simplified atmosphere on top of their structure that insufficiently corrects for the dense winds. 
Without a more sophisticated post-processing, the best way to compare our results is to take the effective temperature of the WN4b stars at the critical point $T_\text{crit}$ as this temperature essentially reflects the outmost radius that could be associated with a hydrostatic regime. 
This value is usually close to the effective temperature of the inner boundary $T_\ast$, but as $T_\ast$ depends on the arbitrary choice of an associated (continuum) optical depth we prefer to use the more physical $T_\text{crit}$.

\subsubsection{Tracks without initial rotation}

\begin{figure}[th]
	\centering
    \includegraphics[width=\hsize]{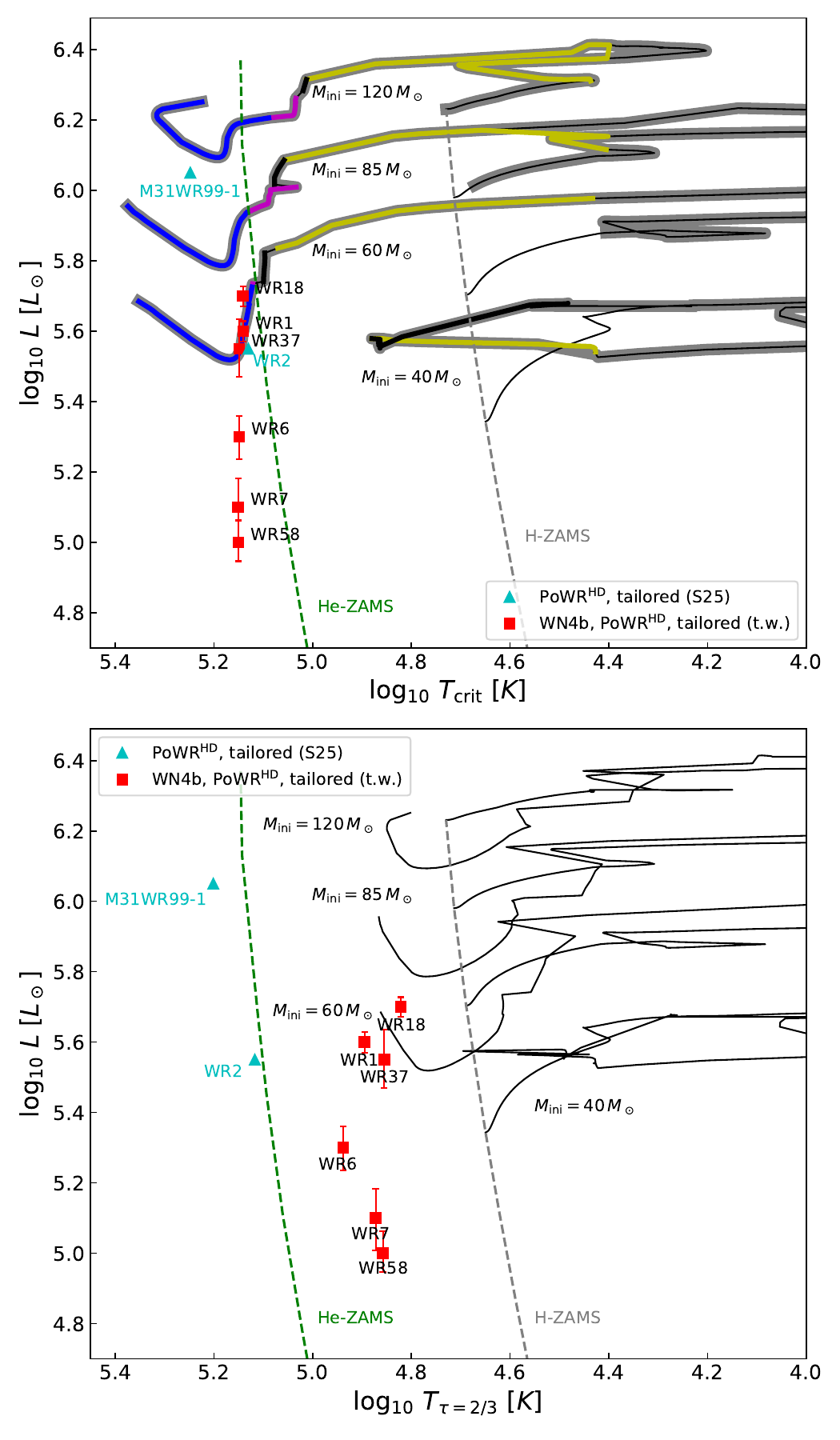}
	\caption{HRD of the PoWR$^\textsc{hd}$ WN4b models (red squares), with the effective temperatures at the critical radius $T_\mathrm{crit}$ (top panel). 
    The GENEC evolutionary tracks without initial rotation from \citet{ekstroem2012} with $M_\mathrm{ini} = 40$, 60, 85 and $120\,M_\odot$ are shown, where we display effective temperatures not corrected for the wind. 
    Separate segments of the evolutionary tracks are colour coded by surface abundance: bold yellow denotes WNh stars ($0.4 > X_\text{H} > 0.02$); bold black denotes WN stars ($X_\text{H} < 0.02$); bold magenta denotes WN/WC stars ($X_\text{C} > 0.02$); and bold blue denotes WC or WO stars ($X_\text{C} > 0.2$). 
    The tracks are further indicated in bold grey according to the criterion $\tau(R_\ast) \gtrsim 1.45$ from \citet{aguilera-dena2022} (see also Eq.\,\eqref{eq:tau_wr}).
    For comparison, the PoWR$^\textsc{hd}$ models of M31WR99-1 and WR2 from \citetalias{sander2025} are shown as well.
    The bottom panel shows the same PoWR$^\textsc{hd}$ models, but now comparing the wind-corrected temperatures given by GENEC with the determined $T_{2/3}$ temperatures at $\tau=2/3$.
    In both panels the H-ZAMS from \citet{ekstroem2012} and the He-ZAMS from \citet{langer1989} are shown, respectively the grey and green dashed lines.}
	\label{fig:evo_nonrot}
\end{figure}

In Fig.\,\ref{fig:evo_nonrot}, we compare our derived HRD locations with the sample of GENEC tracks from \citet{ekstroem2012} which do not consider initial rotation. 
We show the tracks with $M_\mathrm{ini}=40$, 60, 85 and 120 $M_\odot$. 
In addition, the HRD of Fig. \ref{fig:hrd_t23} is shown where the GENEC tracks are displayed with wind-corrected effective temperatures. 
For the WR stages, the GENEC models offer a distinct information of the effective temperatures without their otherwise inherent wind correction. 
We see that in principle our stars are now in a temperature regime where hydrogen-free WR stars are also predicted. 
However, the models with $M_\mathrm{ini}<40M_\odot$ do not evolve to the position in the HRD where WR stars are found and we therefore do not display these tracks. 
This means there are no tracks in this set which reach the derived HRD positions of WR6, WR7, and WR58. 
This remains true for the wind-corrected tracks in the bottom panel of Fig. \ref{fig:evo_nonrot}.
For WR\,1, WR\,18, and WR\,37, there are suitable tracks leading to the obtained HRD positions, but they already predict WC surface abundances in this stage. 
Inspecting the corresponding masses from the track points, we get $\sim$$18.7\,M_\odot$ from the $M_\mathrm{ini}=60\,M_\odot$ track for WR\,18. 
For WR\,37, the closest point is also on the $M_\mathrm{ini}=60\,M_\odot$ track with a mass of $\sim$$14\, M_\odot$ and for WR\,1 this same track gives us $\sim$$15.6\,M_\odot$. 
Compared to the masses we obtain with the spectral analysis ($29.3\,M_\odot$, $14.5\,M_\odot$, and $19.6\,M_\odot$ for WR\,18, WR\,37, and WR\,1, respectively), the evolutionary masses are systematically lower, albeit to varying degrees. This is the opposite to what is usually obtained for evolutionary masses of O stars.
A recent comparison of the 1D PoWR$^\textsc{hd}$ models with 3D RHD simulations indicates that our approach might overestimate the stellar masses by $\sim 10\,\%$ \citep{gonzalez-tora+2025}, but the difference is considerably larger than obtained in these benchmark calculations.

Given the predicted WC appearance, we check whether the stripping during the WR stage itself in the GENEC model is too high. 
The $60\,M_\odot$-track reaches the He-ZAMS with $\sim$$25\,M_\odot$ and shows WC surface abundances approximately 70\,000 years later with $\sim$$20\,M_\odot$ left. 
This corresponds to an average mass-loss rate of $10^{-4.16}\,M_\odot\,\mathrm{yr}^{-1}$, which is about a factor of four to five higher than what we obtain for the three more luminous WN4b stars. 
Considering the individual points of the stars, the mass-loss rates prescribed in the tracks are still typically a factor of $2.5$ higher than we obtain. 
This implies that the time span of the hydrogen-free WN stage is likely underestimated in the tracks and that the predicted spectroscopic WN appearance is more extreme (i.e., stronger emission, cooler appearance) than empirically found. 
This likely also has impacts in the resulting kinetic feedback \citep[e.g.,][]{vieu2024,hawcroft2025,larkin2025} although studying one single WN subtype is too limited to draw broader conclusions. 
Moreover, despite the reasonable approximation of the \citet{nugis_lamers2000} description in the $\dot{M}$-$L/M$-plane, the $L$-dependent implementation, not just in GENEC but in stellar evolution tracks in general, seems to be problematic, even at solar metallicity. 
These specific conclusions of the inherent mass-loss of hydrogen-free WNs on the He-ZAMS are largely independent of how the stars reached this HRD region and thus also remain valid for other stellar evolution codes and binary evolution calculations yielding WR stars. 

\subsubsection{Tracks with initial rotation}

\begin{figure}[h]
	\centering
    \includegraphics[width=\hsize]{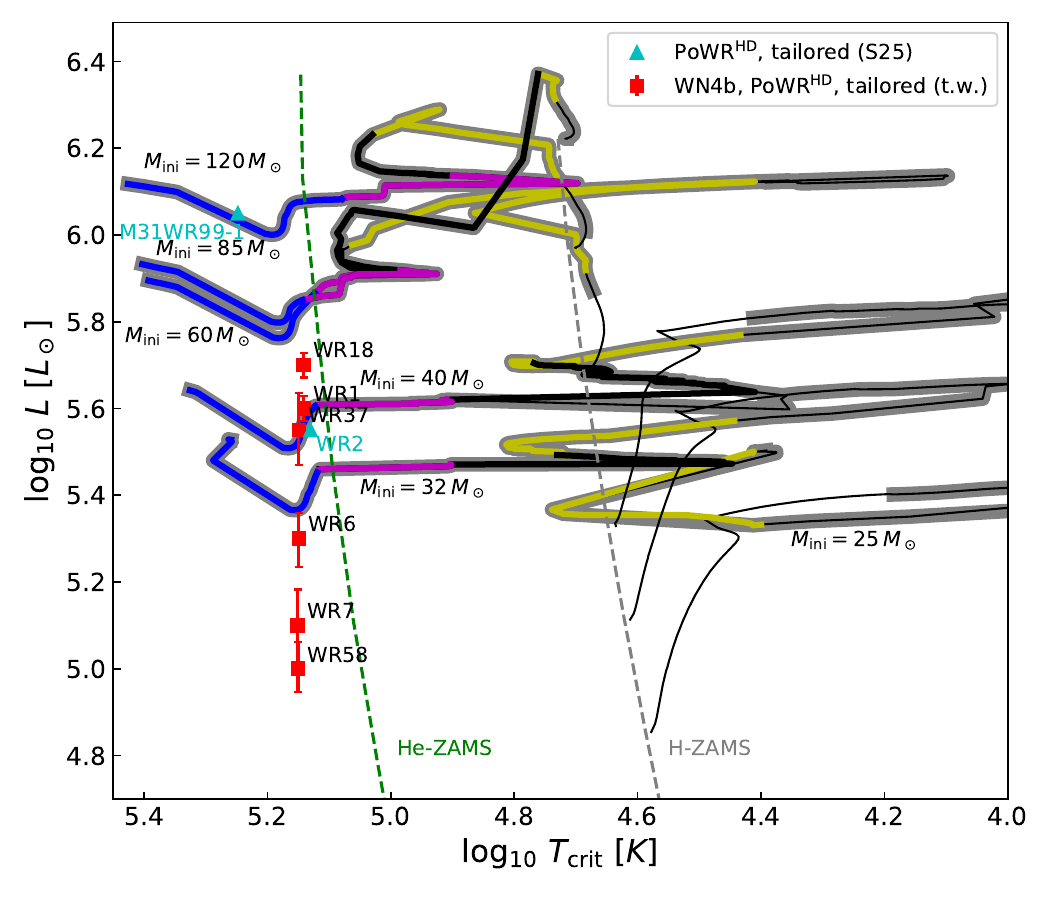}
	\caption{Similar to Fig. \ref{fig:evo_nonrot}, but now with GENEC evolutionary tracks with an initial rotation of $\varv_\mathrm{ini} = 0.4\,\varv_\mathrm{crit}$ \citep{ekstroem2012}.
    Here, the evolutionary tracks with $M_\mathrm{ini}=32$ and $25\,M_\odot$ are shown in addition.}
	\label{fig:evo_rot}
\end{figure}

In Fig.\,\ref{fig:evo_rot}, we consider the GENEC evolutionary models of stars which have an initial rotation velocity of 40\% of the critical rotation, i.e., $\varv_\mathrm{rot,ini} = 0.4\,\varv_\mathrm{crit}$.
Here, the picture shifts somewhat as models having a lower $M_\mathrm{ini}$ can become WR stars, however they still fail to reach parameters similar to WR\,7 and WR\,58. 
WR\,18 falls in between the $M_\mathrm{ini} = 40\,M_\odot$ and $M_\mathrm{ini} = 60\,M_\odot$ tracks here, where the closest points on the tracks respectively give $\sim$$15.7\,M_\odot$ and $\sim$$19.0\,M_\odot$.
For WR\,1 and WR\,37 on the $M_\mathrm{ini}=40\,M_\odot$ track, we get $\sim$$15.5\,M_\odot$ and $\sim$$14.1\,M_\odot$, respectively. 
With the rotating evolutionary models, the track with $M_\mathrm{ini}=32\,M_\odot$ and the point with mass $\sim$$10.5\,M_\odot$ can be representative for WR\,6 (having a spectroscopic mass of $10.3\,M_\odot$).
However, besides WR\,6, the spectroscopic masses still tend to be significantly larger than the evolutionary ones from the tracks with initial rotation. 
Similarly, the discrepancy between the empirically obtained and the observed mass-loss rate remains. 
This is unsurprising as the implemented formula from \citet{nugis_lamers2000} depends mainly on L and thus yields similar values for the same HRD position. 
Again, the match for WR\,6 is the closest with a difference of $0.05\,$dex in $\log L$ and $\sim$$0.15$ dex in $\dot{M}$, albeit with a clear surface abundance mismatch.

\begin{figure}[th]
	\centering
    \includegraphics[width=\hsize]{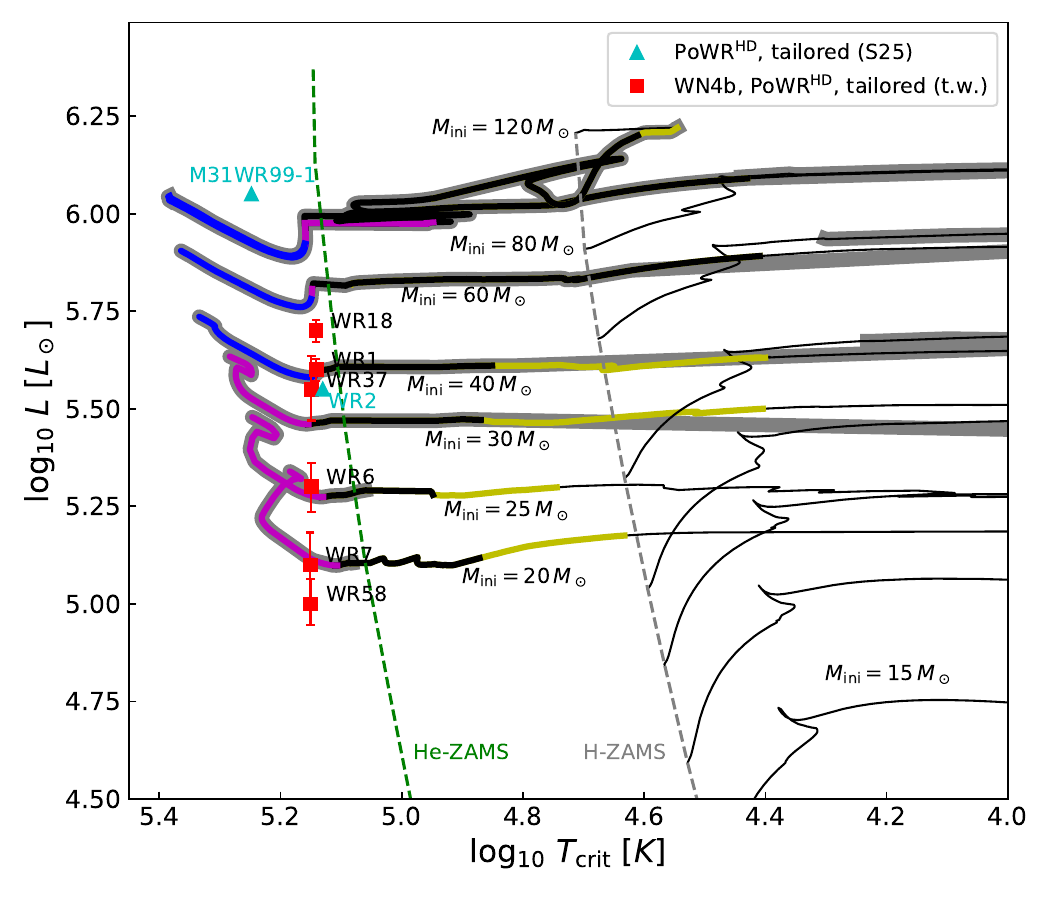}
	\caption{Similar to Fig. \ref{fig:evo_rot}, but now with FRANEC evolutionary tracks \citep{chieffi2013} where the initial rotation has a value of $\varv_\mathrm{ini} = 300\,\mathrm{km}\,\mathrm{s}^{-1}$.}
	\label{fig:evo_rot_franec}
\end{figure}

As an alternative to the GENEC models, we also consider the FRANEC evolutionary tracks from \citet{chieffi2013}, where we make a similar comparison as with the GENEC tracks in Fig. \ref{fig:evo_rot_franec}. 
While the core principles of the codes are not as different as for example GENEC and MESA, the detailed treatment is sufficiently different to considerably affect the predicted evolution paths. 
The FRANEC models assume an initial rotational velocity of $300\,\mathrm{km}\,\mathrm{s}^{-1}$ for all calculations rather than the fixed fraction of critical rotation assumed in GENEC. 
Moreover, the chemical mixing is less efficient \citep[cf.\ Fig.\,11 in][]{chieffi2013}. 
Interestingly, the wind mass-loss treatment is roughly similar to GENEC with the main differences being in the exact treatment of the cool-star wind regime. 
The most noticeable difference between the GENEC and the FRANEC tracks is that the latter reach the He-ZAMS from initial masses as low as $20\,M_\odot$. 
Thus, these tracks reach the positions of WR\,6 and WR\,7, predicting masses of $9.4$ and $7.5\,M_\odot$, respectively. 
This is within the uncertainty range of our obtained masses. For WR\,1, WR\,18 and WR\,37, the tracks predict $\sim$$15.7$, $\sim$$20.0$ and $\sim$$14.9\,M_\odot$, very similar to the GENEC results. 
The predicted current mass-loss rates, however, are notably different. 
For WR\,1, the predicted value of $-4.9$ is in line with the derived one, while for the much less luminous WR\,7, the predicted value of $-5.5$ is a factor of two too low. 
WR\,6, which is intermediate in luminosity to the other two, is predicted to have $-5.34$, which is $0.24\,$dex too low. 
Again, also the surface abundances are off as all targets are predicted to have significant carbon surface abundances. 
While the WN/WC star WR\,58 is not matched by the tracks, the prediction for the slightly more luminous WR\,7 is already $X_\text{C} \approx 0.13$ and $X_\text{O} \approx 0.05$, indicating too strong stripping of the WR star.

Given that both FRANEC and GENEC employ \citet{nugis_lamers2000} for the WR stage, it is puzzling that they differ considerably in their $\dot{M}$ predictions. 
As their initial metallicity is very similar, the reason has to be a different implementation of the recipe. 
In GENEC, the original interpretation of \citet{nugis_lamers2000} is used, where the whole sum of metals ($Z$) enters the calculation. 
In FRANEC, this might not be the case and just the initial metallicity is used. 
Then, the inherent, but not really physical $Y$-dependence \citep[see, e.g., the discussion in][]{higgins2021} would lead to a reduction in $\dot{M}$ if the tracks predict a C-enriched surface (i.e., $Y \lesssim 0.9$), contrary to our findings ($Y \approx 0.98$). 
Using Eq.\,(22) from \citet{nugis_lamers2000} with this assumption approximately yields the FRANEC track values. 
However, the luminosity-dependent discrepancies remain. 
Moreover, while not scaling WR mass-loss rates with carbon is motivated by the findings of \citet{VinkdeKoter2005} that even WC mass-loss rates mainly scale with iron, keeping the He-dependent term without changes is still problematic as it does not reflect the underlying physical effect, which is the difference in the free electron budget \citep{Sander+2020}. 
Hence, it makes a difference whether He is lower due to H or C, but this is not mapped in the augmented \citet{nugis_lamers2000} recipe.

Despite the success to reach the obtained HRD positions, significant caveats thus remain in explaining the evolutionary paths of WR stars. 
The agreement in mass-loss rates for targets such as WR\,1 is due to coincidence rather than coherent treatment. 
Moreover, similar to the GENEC tracks, a high initial rotational velocity is required to have stars eventually reaching the hot side of the HRD in the FRANEC tracks. 
However, rotational velocities of around $300\,\mathrm{km}\,\mathrm{s}^{-1}$ are rarely observed in the O-star regime. 
\citet{Holgado+2022} determined rotational velocities of 285 O-stars which are either single or dominated by the light of one component. 
Their obtained velocity distribution peaks at $\varv \sin i < 100\,\mathrm{km\,s}^{-1}$. 
Even when considering inclination effects, such stars would not rotate fast enough to be described by the FRANEC or GENEC tracks with rotation. 
While smaller secondary peaks exist around $200$ and $300\,\mathrm{km\,s}^{-1}$ in \citet{Holgado+2022}, they are much smaller and seem to correlate more with binary-interaction products than objects undergoing unperturbed main sequence evolution. 
Therefore, the pathways to WR formation, in particular at the lower luminosity end, remain highly uncertain. 
Multiplicity might play an important role in this process, but also needs to explain why none of our sample stars show a luminous companion despite not being runaway stars. 
Compact companions are also unlikely as at least some WN4b stars such as WR\,6 have been extensively monitored over decades.

\subsubsection{Wolf-Rayet identification}

Traditionally, the WR phase and their different subtypes are identified by the surface abundances in stellar evolution models \citep[e.g.,][]{maeder1991,langer1994,chen2015}. 
We follow a similar scheme in the colouring of the tracks in Figs.\,\ref{fig:evo_nonrot} and \ref{fig:evo_rot}. 
However, while the surface abundances might give a reasonable idea of whether we have a WNh, WN or WC/WO type, the abundances as such are not directly related to whether a star is sufficiently close to the Eddington Limit to have an emission-line spectrum and thus appear as a WR star \citep[see, e.g.,][]{Sander+2020,shenar2020}. 
We thus also inspect the criterion from \citet{langer1989} to estimate the wind optical depth via
\begin{equation}\label{eq:tau_wr}
    \tau(R_\ast) = \frac{\kappa_\mathrm{e} \dot{M}}{4\pi R_\ast (\varv_\infty - \varv_0)}\,\ln\left(\frac{\varv_\infty}{\varv_0}\right),
\end{equation}
where $\kappa_\mathrm{e} = 0.2\,(1 - X_\mathrm{H}\,\mathrm{cm}^2\,\mathrm{g}^{-1})$ is the electron scattering opacity and $\varv_0$ is a velocity at the wind onset \citep[Note that $\beta=1$ is assumed to derive the above equation in][]{langer1989}.
To obtain $\varv_\infty$ for the evolution models, we use the expression $\varv_\infty = \alpha\,\varv_\mathrm{esc} = \alpha \sqrt{\frac{2GM}{R}(1 - \Gamma_\text{e})}$ (with $\Gamma_\text{e}$ denoting the Eddington factor for pure electron scattering) and assuming $\alpha \approx 1.3$ \citep[see][]{graefener2017}.
An optically thick wind can generally be defined by the wind optical depth surpassing unity. 
However, a proper definition would include the flux-weighted opacity including all contributions, in particular from line driving. 
To account for the simplifying assumptions in Eq.\,\eqref{eq:tau_wr}, \citet{aguilera-dena2022} attempted an observation-driven calibration, finding $\tau(R_\ast)\gtrsim 1.45$.
Notably, this value is quite high, in particular as the electron scattering is only a fraction of the total opacity. 
\citet{pauli2023} indeed only find $\tau(R_\ast) \gtrsim 0.2$ in order to match observations.
Applying this $\tau(R_\ast)\gtrsim 1.45$ value to the evolutionary tracks in Figs.\,\ref{fig:evo_nonrot}, \ref{fig:evo_rot}, and \ref{fig:evo_rot_franec}, while assuming $\varv_0\approx 20\,\mathrm{km}\,\mathrm{s}^{-1}$, would essentially render the whole tracks as WR stars, which is clearly incorrect. 
In the above figures, we thus use the more conservative value of $\tau(R_\ast) \gtrsim 1.45$. 
Still, the outcome is that large parts of the depicted evolutionary tracks fulfill this criterion, while, with the exception of the very massive star (VMS) regime, this is neither observed nor intended from the calculations. 
For the switch to the VMS mass-loss regime, \citet{sabhahit2022,sabhahit2023} suggest a formalism utilizing the connection between the wind optical depth and the wind efficiency $\eta = \dot{M}\varv_\infty / (L/c)$. 
However, when calculating the wind efficiency for the tracks using the same, conservative assumption for $\varv_\infty$, also almost all shown tracks from the main sequence on yield $\eta > 1$. 
Below the VMS regime, the main sequence should not fulfill this criterion, which is even above of what is actually required to switch on enhanced mass loss in \citet{sabhahit2023}.
Hence, our tests essentially confirm that the mass-loss rates in the OB main sequence regime are overestimated in the applied GENEC tracks.

\section{Conclusions}\label{sec:conclusions}

By using hydrodynamically consistent atmosphere modelling with the PoWR$^\textsc{hd}$ branch, we successfully reproduced observed spectra of six Galactic WN4b stars and derived associated stellar and wind parameters, in combination with using updated Gaia DR3 parallaxes.
Comparing the results of the models in this study with previous analyses in \citetalias{hamann2019} \& \citetalias{sander2019} that used grid models with a prescribed $\beta=1$ velocity-law, we get substantial differences, even after correcting for influences from distance estimates (e.g., on $L_\ast$, $R_\ast$ and $\dot{M}$).

With the PoWR$^\textsc{hd}$ modelling approach, we reach a similar quality with synthetic spectrum, while also breaking previous degeneracies in the spectral analysis of our sample stars. 
Due to the solution of the wind hydrodynamics, the observed spectrum and its necessary mass-loss rate can only be reached with a much more compact solution than obtained in $\beta$-type modelling. 
All six WN4b targets have very similar effective temperatures at the critical radius, $T_\mathrm{crit}\sim\,140$ kK; a value also representative for the stellar radius $T_\ast$ as the difference is quite small in our models (see Table\,\ref{tab:stellar_params}). 
In contrast, prior analyses had the sample spread between $T_\ast \approx 79$ and $112\,$kK, though with the potential to also provide hotter solutions due to the spectrum fully emerging in the dense wind.

With their updated temperatures, the explored WN4 stars are now close or slightly hotter than the He-ZAMS. 
As none of our targets show signatures of hydrogen, such a location is expected from stellar structure modelling and we thus solve the long-standing ``WR radius problem'', at least for the WN4b subtype. 
This finding aligns with the prototypical study by \citet{graefener2005} for the WC star WR\,111 and the recent WN2 analysis by \citet{sander2025}, both also employing hydrodynamically-consistent models. 
In all of these cases, the stellar wind is launched deeply at the hot iron bump created mainly by Fe M-shell opacities.

When comparing the updated mass-loss rates $\dot{M}$ with current predictions, we see that the \citet{SanderVink2020} $\dot{M}(L/M)$-description systematically underpredicts $\dot{M}$, while the transformed mass-loss rates $\dot{M}_\mathrm{t}$ align well with predicted relations for the thick wind regime. 
As both works use essentially the same atmosphere modelling methodology, a major reason for this discrepancy is a difference between the mass-luminosity-relation inherent to \citet{SanderVink2020} compared to what we found for our WN4b sample.
As the same or similar relations were used in earlier studies to infer masses of WR stars, we conclude that stellar mass estimates from He-ZAMS-based relations such as \citet{graefener2011} will usually be too high, at least for strong-wind, early-type WR stars. 
This is notably different for the weaker-winded WN2 stars in \citet{sander2025}, which seem to closely align with a He-ZAMS relation.

When considering the mass-loss descriptions from \citet{nugis_lamers2000} and \citet{yoon2017}, which have an $\dot{M}(L)$-type description, we conclude that both recipes perform well on average for the WN4b stars at $Z \approx 0.014$. 
However, both recipes fail to reproduce the WN2 stars from \citet{sander2025}, while \citet{SanderVink2020} does a reasonable job here and potentially could even offer an indirect metallicity diagnostic when the ``breakdown regime'' in $\dot{M}$ \citep[cf.\,][]{SanderVink2020} is empirically mapped with WN2-like objects. 
Thus, we have to conclude that empirical recipes perform well if applied within the typical regime they were designed for, but have problems to cover more rare objects and suitably deal with lower metallicity.

Secondly, both GENEC and FRANEC models predict our WN4b stars to have a carbon-rich (i.e, WC- or WO-like) surface, which we can clearly rule out from our spectroscopic analysis. 
One of our objects, WR\,58, is of WN/WC-transition type, but the derived carbon abundance is way below the level of 1\% typically expected for these transitions types. 
The inherent, unphysical dependencies on the current He- and $Z$-abundance in the \citet{nugis_lamers2000} description lead to an overprediction of mass loss during the WR stage in the GENEC models. FRANEC seems to augment the recipe, resulting in an underprediction for lower luminosities. 
In cases where evolutionary tracks reach the derived HRD positions, there is a systematic difference on the stellar masses $M_\ast$: for higher-luminosity objects, the mass derived from the PoWR$^\textsc{hd}$ modelling is higher than the mass from the tracks. 
For the lower-luminosity objects, the match is better, but tracks with a likely too high initial rotation need to be invoked.

The broader implications of our analysis and conclusions will have to be tested in future studies involving dense-wind Wolf-Rayet stars at different metallicity (e.g., in the LMC) as well as exploring more subtypes and wind density regimes. 
The solution of the ``WR radius problem'' for WN4b and WN2 stars is possible with the current modelling framework as both types of objects can be explained with winds that launch deep enough to not trigger larger amounts of radiation-driven turbulence. 
This is not the case for many other WR subtype configurations such as the weak-lined WN3 stars at lower metallicity \citep[e.g.,][]{gonzalez-tora2025_lvm} which are located in a regime that cannot be explained by deep wind launching \citep[e.g.,][]{grassitelli2018,sander2023temp} and will be significantly affected by turbulent pressure found in 3D and 2D simulations \citep{moens2022,moens2025}. 
To account for these, further developments in incorporating multi-D effects in 1D atmosphere modelling will be necessary. 
The mismatch to correctly represent or even reach the observed WR configurations with stellar evolution models further has significant effects on population synthesis, both in terms of stellar feedback as well as spectral predictions, and demands a deeper exploration of possible evolutionary pathways with updated physical treatments.

\section{Data availability}

Appendix B (spectral fits) and figures therein can be found in \url{https://zenodo.org/records/18220388}.

\begin{acknowledgements}
      The authors would like to explicitly thank the anonymous referee for providing helpful comments that improved the quality of this document. 
      The authors also want to thank R.\ Hirschi, G.N.\ Sabhahit, and J.S.\ Vink for inspiring discussions of some preliminary results that helped to outline the scope of this paper.
      
      RRL, AACS and MBP are supported by the Deutsche Forschungsgemeinschaft (DFG, German Research Foundation) in the form of an Emmy Noether Research Group – Project-ID 445674056 (SA4064/1-1, PI Sander).
      GGT and JJ are supported by the Deutsche Forschungsgemeinschaft (DFG, German Research Foundation) under Project-ID 496854903 (SA4064/2-1, PI Sander).
      GGT and AACS further acknowledge funding from the Federal Ministry for Economic Affairs and Climate Action (BMWK) via the Deutsches Zentrum für Luft- und Raumfahrt (DLR) grant 50 OR 2503 (PI Sander).
      VR acknowledges financial support by the Federal Ministry for Economic Affairs and Energy (BMWE) via the Deutsches Zentrum f\"ur Luft- und Raumfahrt (DLR) grant 50 OR 2509 (PI Sander).
      AACS, VR, and ECS further acknowledge financial support by the Federal Ministry for Economic Affairs and Climate Action (BMWK) via the German Aerospace Center (Deutsches Zentrum f\"ur Luft- und Raumfahrt, DLR) grant 50 OR 2306 (PI: Ramachandran/Sander).
      RRL, MBP, JJ and ECS are members of the International Max Planck Research School for Astronomy and Cosmic Physics at the University of Heidelberg (IMPRS-HD).
      This project was co-funded by the European Union (Project 101183150 - OCEANS).
\end{acknowledgements}

\nocite{*}
\bibliographystyle{aa}
\bibliography{paperbib.bib}

\begin{appendix}\label{appendix}
\onecolumn

\section{ionising fluxes, abundances, and modelled elemental ions}

\begin{figure}[h]
	\centering
    \includegraphics[width=\hsize]{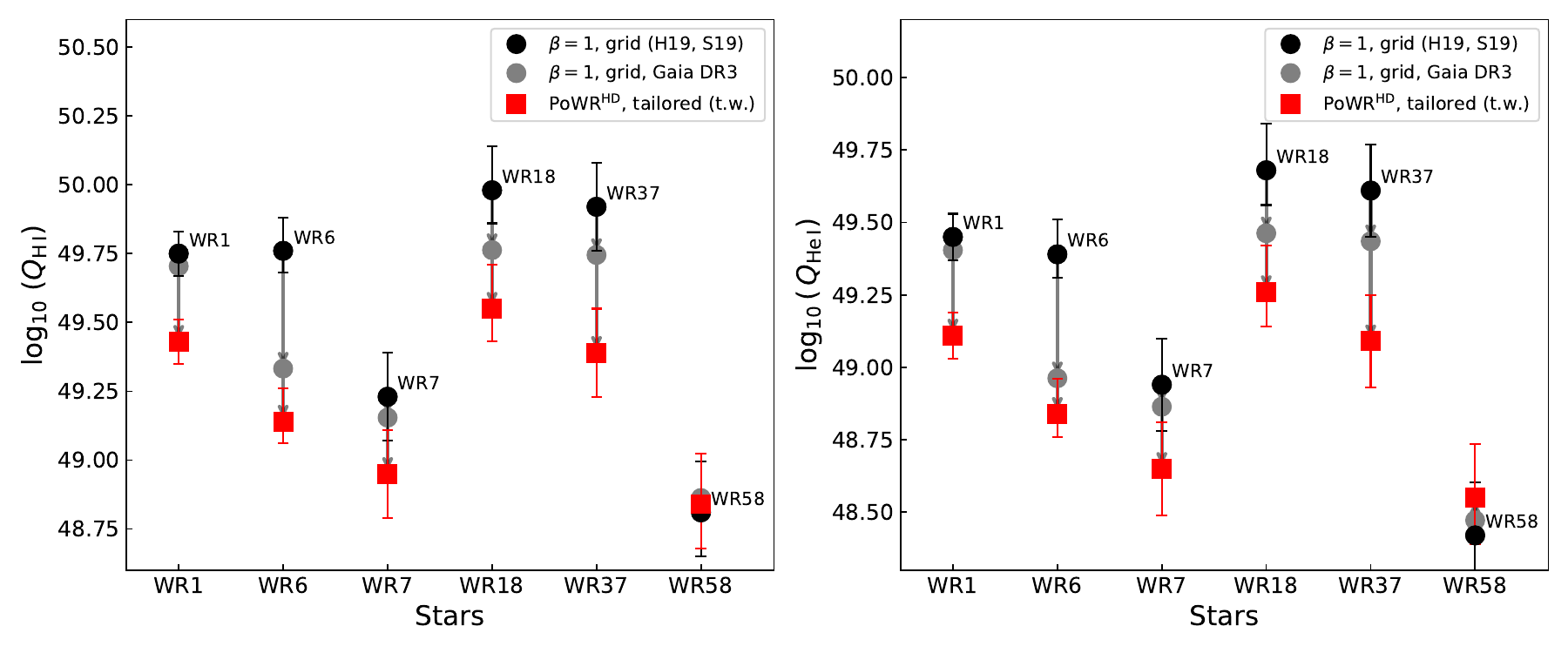}
	\caption{The amount of ionising photons for \ion{H}{i} ($Q_\ion{H}{i}$, left panel) and for \ion{He}{i} ($Q_\ion{H}{i}$, right panel) in case of the $\beta=1$ grid models (\citetalias{hamann2019} \& \citetalias{sander2019}, black circles) and the PoWR$^\textsc{hd}$ models (red squares). 
    Similar to Figs. \ref{fig:hrd}, \ref{fig:hrd_t23}, and \ref{fig:mdot_comp}, a correction for GAIA DR3 parallaxes (grey circles) is made in order to separate the effects of distance updates and model differences, denoted by the grey arrows.}
	\label{fig:qhe_comparison}
\end{figure}

\begin{table}[h]
    \caption{Elements, mass fractions $X_\mathrm{m}$ and ions used for the PoWR$^\textsc{hd}$ models in this work$^\mathrm{a}$.}
    \centering
    \def\arraystretch{1.3}
    \label{tab:abundances}
    \begin{tabular}{lll}
        \hline
        Element & $X_\mathrm{m}$      & Ions \\
        \hline
        He      & (0.99)               & I, II, III \\       
        C$^\mathrm{b}$ & ($5.0\cdot 10^{-5}$ for WR1) & I, II, III, IV, V, VI \\
                & ($1.0\cdot 10^{-5}$ for WR6, WR7, WR37) & \\
                & ($2.0\cdot 10^{-5}$ for WR18) & \\
                & ($4.0\cdot 10^{-4}$ for WR58) & \\
        N$^\mathrm{b}$ & ($1.5\cdot 10^{-2}$ for WR7) & I, II, III, IV, V \\
                & ($1.0\cdot 10^{-2}$ for rest) & \\ 
        O       & ($1.0\cdot 10^{-4}$) & I, II, III, IV, V, VI \\
        Ne      & ($1.3\cdot 10^{-3}$) & I, II, III, IV, V, VI, VII \\
        Na      & ($2.7\cdot 10^{-5}$) & I, II, III, IV, V \\
        Mg      & ($6.9\cdot 10^{-4}$) & I, II, III, IV \\         
        Al      & ($5.3\cdot 10^{-5}$) & I, II, III, IV, V \\
        Si      & ($6.7\cdot 10^{-4}$) & I, II, III, IV, V, VI \\
        P       & ($5.8\cdot 10^{-6}$) & II, III, IV, V, VI \\
        S       & ($3.1\cdot 10^{-4}$) & I, II, III, IV, V, VI, VII \\
        Cl      & ($8.2\cdot 10^{-6}$) & III, IV, V, VI, VII \\
        Ar      & ($7.3\cdot 10^{-5}$) & I, II, III, IV, V, VI, VII, VIII \\
        K       & ($3.1\cdot 10^{-6}$) & I, II, III, IV, V, VI, VII \\
        Ca      & ($6.1\cdot 10^{-5}$) & II, III, IV, V, VI, VII, VIII \\
        Fe      & ($1.6\cdot 10^{-3}$) & II, III, IV, V, VI, VII \\
        \hline
    \end{tabular}
    \tablefoot{
        \tablefoottext{a}{C, N and O abundances are taken from typical values for WN-stars, values also adopted in the public PoWR WN grids \citep{todt2015}. 
        The remainder of the element abundances are taken from the solar values derived in \citet{asplund2009} and \citet{Scott2015, Scott2015b}.}
        \tablefoottext{b}{Depending on the star, adaptations to C and N abundances are made in order to improve spectral fits.}
    }
    \label{tab:elements}
\end{table}

\end{appendix}

\end{document}